\documentclass[11pt,psfig]{article}
\usepackage{psfig}
\title{Flexion and Skewness in Map Projections of the Earth}
\author{David M. Goldberg (Drexel University Dept. of Physics)\\ J. Richard Gott III
  (Princeton University Dept. of Astrophysical Sciences)\\
{\it goldberg@drexel.edu}, {\it jrg@astro.princeton.edu}}
\begin{document}
\def\ol{\overline}
\maketitle

\begin{abstract}
 Tissot indicatrices have provided visual measures of local area and
isotropy distortions.  Here we show how large scale distortions of
flexion (bending) and skewness (lopsidedness) can be measured.  Area
and isotropy distortions depend on the map projection metric, flexion
and skewness, which manifest themselves on continental scales, depend
on the first derivatives of the metric.  We introduce new indicatrices
that show not only area and isotropy distortions but flexion and
skewness as well.  We present a table showing error measures for area,
isotropy, flexion, skewness, distances, and boundary cuts allowing us
to compare different world map projections.  We find that the
Winkel-Tripel projection (already adopted for world maps by the
National Geographic), has low distortion on most measures and
excellent quality overall.
\end{abstract}


\section{Introduction}

Tissot (1881) indicatrices have been very useful for providing a visual
presentation of local distortions in map projections in a simple and
compelling fashion.  A small circle of tiny radius (say 0.1 degree of
arc in radius) is constructed at a given location, and then enlarged
and projected on the map at that location.  This always produces an
ellipse.  Usually one favors conformal map projections that minimize
the changes in scale factor, or equal area projections that minimize
anisotropy, or recently, map projections that are neither conformal
nor equal area, but which have a judicious combination of minimizing
both scale and isotropy errors (like the Winkel-tripel used by the
National Geographic Society for world maps).

The Tissot ellipse at a given location is specified by three
parameters, the major axis, the minor axis, and the orientation angle
$\theta$ of the major axis of the ellipse.  Geometrically, from
differential geometry (and General Relativity) we know that the
measurement of local distances is measured by the metric tensor
$g_{ab}$, where a and b can each take the values x or y, yielding
three independent components.  Locally for two nearby points
separated by infinitesimal map coordinate differences dx and dy, the
true distance between these two points on the globe is given by:
\begin{equation}
ds^2=g_{xx}dx^2+2 g_{xy}dx\ dy +g_{yy}dy^2
\end{equation}
The Tissot ellipse (major axis, minor axis, and orientation angle,
$\theta$) can be calculated from the three components of the metric
tensor.  Thus, the Tissot ellipse essentially carries the information
on the metric tensor for the map.  It tells us how local infinitesimal
distances on the map correlate with local infinitesimal distances on
the globe.

As we will see, however, the Tissot ellipse does not carry all of the
information related to distortions, which has been noted previously.
Others (Stewart 1943; Peters 1975, 1978; Albinus 1981; Canters 1989,
2002; Laskowski 1997ab) have noted that there are finite distortions
apart from those of the Tissot ellipse.

Previous authors have previously depicted finite size distortions by
several methods, including showing faces on a globe (Reeves 1910;
Gedymin 1946), 30
degree x 30 degree equiangular quadralaterals (Chamberlin 1947), a net of
twenty spherical triangles, in an icosahedral arrangement (Fisher \&
Miller 1944), 150 degree great circle arcs, and circles of 15 degrees radius
Tobler (1964). 

In the Oxford Hammond Atlas of the World (1993) , new conformal map
projections ("Hammond Optimal Conformal Projections") were designed
for the continents.  Following the Chebyshev criterion (Snyder 1993),
the rms scale factor errors were minimized by producing a constant
scale factor along the boundary of the continent.  (For a circular
region this conformal map would be a stereographic projection.)  By
tailoring the boundary to the shape of each continent, the errors
could be reduced relative to those in a simple stereographic
projection.  In touting the advantages of their projection the Hammond
Atlas did the following experiment.  Following Reeves (Reeves 1910), they
constructed a face on the globe with a triangular nose, a straight
(geodesic) mouth, and eyes that were pairs of concentric circles on
the globe.  They then showed this face with various map projections.
In the gnomonic, the mouth was straight, but the eye circles were not
circular and were not concentric (what we would call skewness).  The
Mercator projection had the mouth smiling (what we would call flexion)
and although the eyes were circular they were not concentric.  The
Hammond Optimal Conformal projections did a bit better on these
qualities because the gradients of the scale factor changes were small
so the flexion and skewness were small, although of course not zero.

In Sec. \ref{sec:flexion} we introduce the concept of ``flexion'', by
which a map projection can cause artificial bending of large
structures.  In Sec. \ref{sec:skewness}, we show another distortion:
skewness, which represents lopsidedness and an asymmetric stretching
of large structures.  We show a simple way to visualize these
distortions in Sec. \ref{sec:gg}, in which we introduce the
``Goldberg-Gott Indicatrices.'' In Sec. \ref{sec:geometry} we derive a
differential geometry approach to measuring flexion and skewness.
While readers interested in computing flexion and skewness on
projections not included in this paper should refer to
Sec. \ref{sec:geometry}, it is highly technical, and those interested
in seeing results may skip directly to Sec. \ref{sec:discussion}, in
which we discuss Monte Carlo estimates of the distortions for a number
of projections, a ranking of map projections, and our conclusions.

\section{First Distortion: Flexion}

\label{sec:flexion}

The local effects shown by the Tissot Ellipse are not the only
distortions present in maps.  There are also ``flexion'' (or bending)
and ``skewness'' (or lopsidedness; discussed in the next section),
which describe curvature distortions visible on world maps (the
terminology stems from a similar effect in gravitational lensing;
Goldberg \& Bacon 2005).

One can think about flexion in the following way.  Imagine a truck
going along a geodesic of the globe at unit angular speed (say one
radian per day).  Now imagine the image of that truck on the map,
moving along.  If the map were perfect, if it had zero flexion and
zero skewness, then that truck would move in a straight line on the
map with constant speed.  Its velocity vector on the map:
\begin{equation}
{\bf v}=\frac{d{\bf x}}{d\tau}\ ,
\end{equation}
would be a constant, where $\tau$ is the angle of arclength in radians
 traveled by the truck along the geodesic on the globe.  Its acceleration:
\begin{equation}
{\bf a}=\frac{d{\bf v}}{d\tau}
\end{equation} 
would be zero.  Of course, this cannot be true for a general geodesic.
In the general case, the image of the truck suffers an acceleration as
it moves along.  The acceleration vector, ${\bf a}$, in the
two-dimensional map has two independent components: $a_\perp$ (which
is perpendicular to the truck's velocity vector at that point), and
$a_\parallel$ (which is parallel to its velocity vector at that
point).

The perpendicular acceleration, $a_\perp$, causes the truck to turn
without changing its speed on the map.  This causes flexion, or
bending, of geodesics.  We define the flexion along a given geodesic
at a given point to be:
\begin{equation}
f=\frac{a_\perp}{v}\ ,
\end{equation}
or, more usefully:
\begin{equation}
f=\frac{dv_\perp}{d\tau}\frac{1}{v}\ .
\end{equation}
In this form, we may define:
\begin{equation}
d \alpha=\frac{dv_\perp}{v}\ .
\end{equation}
where $\alpha$ is the angle of rotation suffered by the velocity
vector.  Remember, if $a_\perp$ is the only acceleration present, then
the velocity vector of the truck on the map does not change in
magnitude, but just rotates in angle.  Thus $f = d\alpha/d\tau$, and
represents the angular rate at which the velocity vector rotates
divided by the angular rate at which the truck moves on the globe.

Skewness and flexion only express themselves on large scales.  They
are not noticeable on infinitesimal scales where the metric contains
all the information one needs, but become noticeable on finite scales,
with their importance growing with the size of the area being examined.
Flexion and skewness are thus important on continental scales and
larger in a world map.

It is possible to design a map projection that has zero flexion: the
gnomonic projection shows all great circles as straight lines.
However it does exhibit anisotropy, scale changes, and skewness, and
at best can show only one hemisphere of the globe.

\subsection{Example 1: The Stereographic Projection}

\begin{figure}[h]
\centerline{\psfig{figure=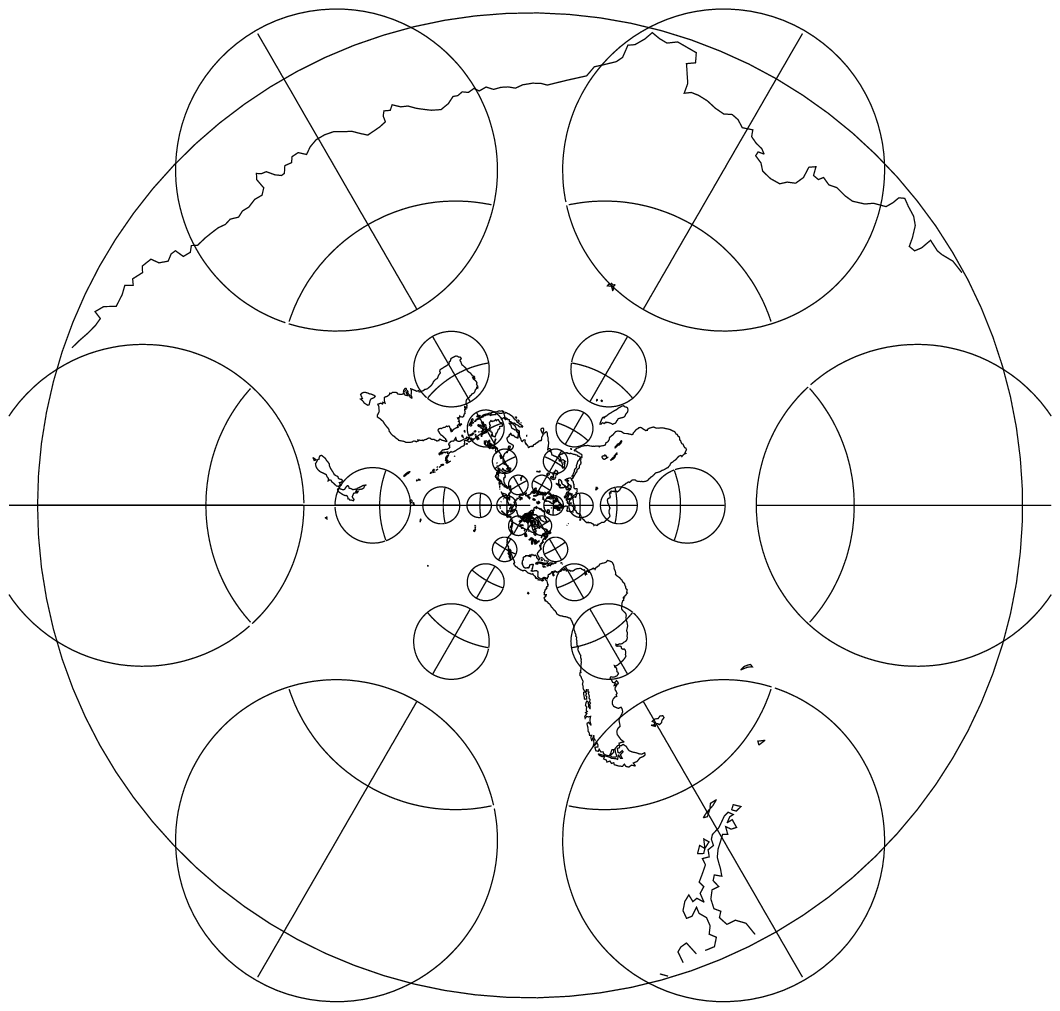,angle=0,height=3.7in}}
\caption{The Indicatrix map for a Stereographic projection.}
\label{fg:stereographic}
\vspace{0.2in}
\centerline{\psfig{figure=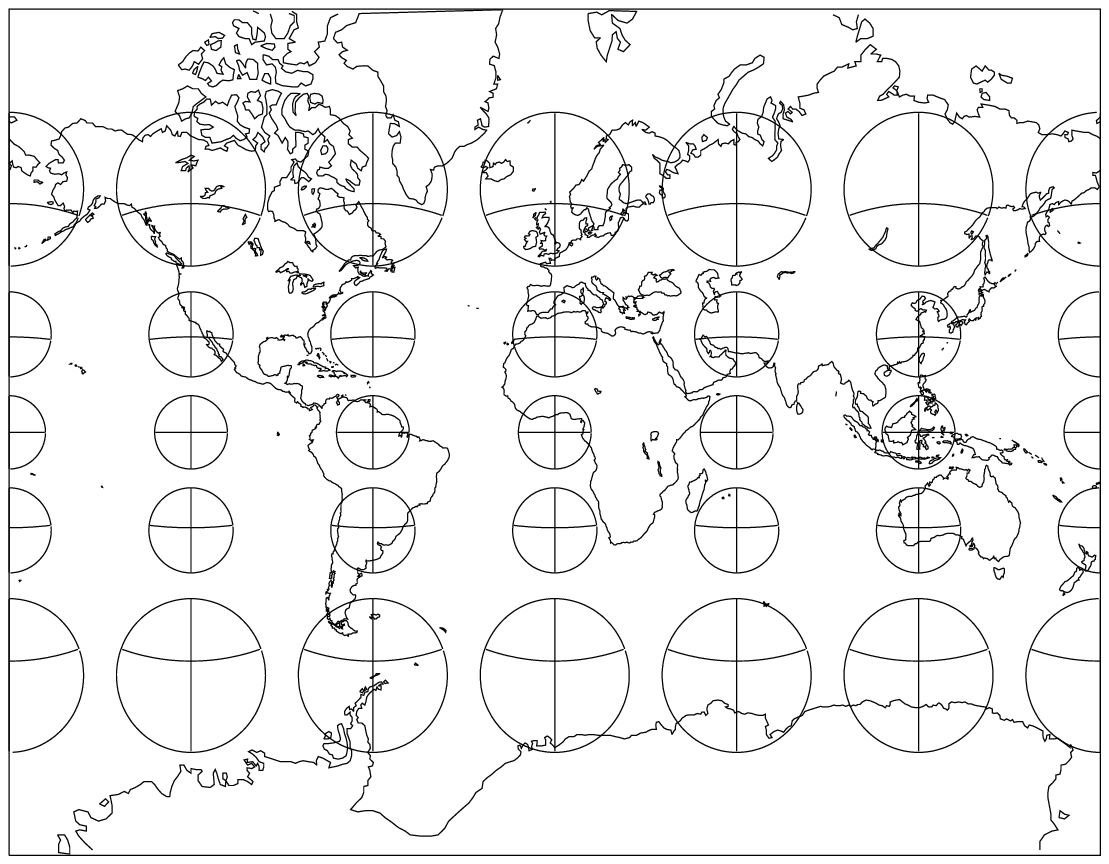,angle=0,height=3.7in}}
\caption{The Indicatrix map for a Mercator projection.}
\label{fg:mercator}
\end{figure}

As an example, consider a truck traveling on the equator as seen by in
the polar stereographic projection (see Fig~\ref{fg:stereographic}).
In the stereographic projection, the north pole is in the center of
the map and the equator is a circle around it.  As the truck circles
the equator (the equator is a geodesic-so the truck drives straight
ahead on the globe), it travels around a circle on the map.  By
azimuthal symmetry, the truck circles the equatorial circle on the map
at a uniform rate.  The velocity vector of the truck on the map
rotates a complete $360^\circ$ ($2\pi$ radians), as the truck circles
the equator, traversing $360^\circ$ of arc on the globe.  So the
flexion is f = 1, for a point on the equator, for a geodesic pointing
in the direction of the equator.  The flexion is defined at a point,
and for a specific geodesic traveling through that point.

For an arbitrary point on the equator in the stereographic projection
and a geodesic pointing in the direction of a meridian of longitude
(also a geodesic) the flexion is zero, because these geodesics are
shown as radial straight lines in the polar stereographic projection,
and the velocity vector of the truck does not turn as it travels
north.

The stereographic projection has the property that every great circle
(geodesic) on the globe is shown as a circle on the map, except for a
set of measure zero that pass through the north pole (i.e. the
meridians of longitude).  Thus by the argument given above, the
average flexion integrated around a random great circle must be
$\langle f\rangle = 1$, because the truck's velocity vector on a
random great circle must rotate by $360^\circ$ as it circles the
$360^\circ$ of arc completing that great circle on the globe.  The
magnitude of the velocity vector of the truck on the map is larger the
further from the north pole it is and so its rotation per angle of arc
of truck travel on the globe is larger there as well, and so the
flexion along that random geodesic is larger the further away from the
pole one is, with the integrated average value along the whole great
circle being $\langle f \rangle =1$.

\subsection{Example 2: The Mercator Projection}

The Mercator projection (see Fig~\ref{fg:mercator}) is conformal and
so only the scale factor changes as a function of position on the map
(i.e. $g_{xx} = g_{yy}$, and $g_{xy} = 0$ and the Tissot ellipses are
all circles with radii proportional to 1/$g_{xx}$).  But there is
bending.  The northern boundary between the continental United States
and Canada at the $49^{th}$ parallel of latitude is shown as a
straight line in the Mercator Map, but really it is a small circle
that is concave to the north.  If one drove a truck down that border
from west to east, one would have to turn the steering wheel slightly
to the left so that one was continually changing direction.  The great
circle route (the straightest route) connecting the Washington State
and Minnesota (both at the 49th parallel) is a straight line which
goes entirely through Canada.  This straight line on the globe when
extended, passes south of the northern part of Maine, so the
continental United States is bend downward like a frown in the
Mercator Map.  (See Figure ~\ref{fg:us_mercator} and
~\ref{fg:us_oblique_mercator}).  Likewise, Maine on a Mercator map
Maine sags below the line connecting Washington State and Minnesota,
while on the globe this is not true.

In the Mercator projection, the flexion along the equator is zero,
also along all meridians of longitude, but these are a set of measure
zero.  A random geodesic is a great circle that is inclined at some
angle between $0^\circ$ and $90^\circ$ with respect to the equator.
On the Mercator map this is a wavy line that bends downward in the
northern hemisphere, and by symmetry, upward in the southern. Since
the curvatures are equal and opposite in the two hemispheres, the
average flexion $\langle f\rangle =0$, but this is misleading because
the flexion at each point off the equator is not zero.  So if we are
rating map projections by the amount of flexion they contain we should
use the absolute value of the flexion instead: $|f|$.  In a region
where the flexion does not change sign (such as the northern
hemisphere in the Mercator projection or the entire stereographic map)
the total bending of a geodesic segment will be the integral of the
flexion $|f|$ over that segment.  In fact, in Section
\ref{sec:numerical} we will evaluate the overall flexion on a map by
simply picking random points on the sphere and random directions for
geodesics going through them, and then calculating the absolute value
for the flexion for all random points on the globe and random
directions through them.

\subsection{A Global Flexion Measure}

We can calculate the flexion for any point in the Mercator (or any
other) projection through any geodesic using spherical trigonometry.
As a reminder to the reader, the Mercator projection uses the mapping:
\begin{eqnarray}
x&=&\lambda \\ \nonumber  
y&=&\ln\left(\tan [\pi/4+\phi/2]\right)\\
\end{eqnarray}
where here and throughout, $\lambda$ is the longitude expressed in
radians, and $\phi$ is the latitude expressed in radians.  

On the map, the angle rotation along the geodesic with an azimuth,
$\theta$, is calculated by constructing thin spherical triangle with side-angle-side given by
$(\pi/2 - \phi, \theta,d\tau)$.  
\begin{equation}
d\alpha=\pi-\beta-\theta
\end{equation}
because the geodesic intersects the north-south meridian (a vertical
line in the Mercator) at an angle of $\theta$ initially, and at an
angle $\pi-\beta$ at the other end, where $\beta$ is the angle in the
spherical triangle at the other end of the $d\tau$ side.  Solving for
$d\alpha$ using spherical trigonometry in the limit as $d\tau$ goes to
zero, we find that
\begin{equation}
f =  sin \theta tan \phi
\end{equation}

Thus, for $\theta = \pi/2$, an east-west geodesic, we find that 
\begin{equation}
f_{EW} = tan \phi
\end{equation}
so that in the northern hemisphere, traveling east one's geodesic is
bending clockwise with $d\alpha/d\tau = tan \phi$.  Therefore, the
east-west geodesic bends downward.  For, $\theta = 0$, a north-south
geodesic, the flexion is zero, as we expect, since the meridians of
longitude are straight in the Mercator map.  If we average over all
azimuths at a given point, we find:
\begin{equation}
\langle |f(\phi)| \rangle = |\tan\phi|\frac{2}{\pi}
\end{equation}

Now we can integrate this over all points on the sphere to produce the
average flexion over the whole sphere F.  Taking advantage of the
symmetry between the northern and southern hemisphere we can integrate
only over the northern hemisphere (where $dA = 2\pi \cos\phi d\phi$),
yielding:
\begin{eqnarray}
F&=&\langle|f|\rangle  \\ \nonumber
&=&\frac{\int 2\pi \cos\phi \tan\phi \frac{2}{\pi} d\phi}{2\pi}\\ \nonumber
&=&\frac{2}{\pi}
\end{eqnarray}

The flexion is less than that of the stereographic because of the
$180^\circ$ boundary cut along the longitude line at the international
date line.  A geodesic is a great circle on the globe, and if this is
shown as a closed curve on the map that is always concave inward (the
best possible case) it will always have a total rotation of the
velocity vector of $360^\circ$ and so will have an average integrated
flexion of 1.  If there is a boundary cut, the great circle does not
have to close on the map (it has two loose ends at the boundary cut)
and so need not completely rotate by $360^\circ$.

\section{Second Distortion: Skewness}

\label{sec:skewness}

Acceleration in the direction parallel to the velocity vector of the
truck $a_\parallel$, causes the truck to increase its speed along the
geodesic curve without causing any rotation.  This causes skewness,
because as the truck accelerates, it covers more distance on the map
on one side of a point than on the other, so the point in question
will not be at the center of the line segment of arc centered on that
point on the sphere.

Consider a segment of a meridian of longitude on the globe centered at
$45^\circ$ north latitude.  Going from South to North along that
geodesic in the Mercator map the truck is accelerating with
$a_\parallel > 0$, because the scale factor is getting larger and
larger the further north one goes, so as the truck continues to cover
equal arc length on the globe it covers larger and larger distances on
the map.  Thus, the center of the segment (at $45^\circ$ latitude) is not
centered on the segment on the map. 

We define the skewness:
\begin{equation}
s\equiv \frac{a_{\parallel}}{v}
\end{equation}

Taking the explicit case of the Mercator projection, we find:
\begin{eqnarray}
v_y&=&\frac{dy}{d\phi}\\ \nonumber
&=&\frac{1}{\cos\phi}
\end{eqnarray}
and thus:
\begin{equation}
a_\parallel=\frac{\tan\phi}{\cos\phi} \ ,
\end{equation}
so the skewness (for a vector pointed N-S) is simply:
\begin{equation}
s=\tan\phi\ .
\end{equation}

The skewness at $45^\circ$ is 1, showing a lopsidedness toward the
north. Given this relation, the skewness is positive (northward lopsidedness) in the northern
hemisphere and negative (southward lopsidedness) in the southern
hemisphere.  

Consider a geodesic through a point in the northern hemisphere tipped
at an azimuth angle of $\theta$ with respect to north.  The only thing
increasing the speed of the truck is the gradient of the scale factor
as one moves northward, so the amplitude of the parallel acceleration
is equal to the maximum acceleration (obtained going straight north)
times cos $\theta$.  To get the average of the absolute value of the skewness
for all geodesics through that point at all random angles $\theta$, one
simply integrates over $\theta$:
\begin{eqnarray}
<|s|>&=&|\tan\phi|\frac{\int_0^{\pi/2}\cos\theta d\theta}{\pi/2}\\
\nonumber
&=&|\tan\phi|\frac{2}{\pi}
\end{eqnarray}
 As with the flexion, we can integrate this over all points on the sphere to produce the
average skewness over the whole sphere S.  Similarly, we find:
\begin{equation}
S=\frac{2}{\pi}
\end{equation}

Notice that this is exactly the same value as the average flexion,
$F$, for the Mercator.  We will find that for conformal projections,
the average absolute value of the skewness and flexion at a given
point and over the whole globe are always equal.  (This is only true
for conformal projections, for general projections the skewness and
flexion can be different, as illustrated by the gnomonic projection
which has zero flexion but non-zero skewness.)

In the Mercator projection, at the equator, the skewness is zero, as
we would expect from symmetry considerations.  Indeed, because any
geodesic crossing equator has a symmetric shape in the northern and
southern hemisphere, the skewness s = 0 for any geodesic line
evaluated at a point on the equator. Likewise, the flexion is zero for
any geodesic line evaluated at a point on the equator.  So the
Mercator map has perfect local shapes along the equator, uniform scale
along the equator (Tissot ellipses all equal size circles) and zero
flexion and skewness along geodesics in any direction from points on
the equator.  

While much of the analysis in this work specifically addresses
distortions in maps of the earth, these effects must also be taken
into account in other maps as well.  One of us (JRG) has recently
produced a conformal ``map of the universe'' (Gott et al. 2005) based on the logarithm
map of the complex plane.  The horizontal coordinate is the celestial
longitude in radians, yielding a $360^\circ$ panorama from left to
right.  The vertical coordinate is:
\begin{equation}
y=\ln(d/r_\oplus)
\end{equation}
where $d$ is the distance, and $r_\oplus$ is the radius of the earth.
The distance scale goes inversely as the distance, allowing us to plot
everything from satellites in low earth orbit to stars and galaxies,
to the cosmic microwave background on one map.  The map is conformal,
having perfect local shapes.  However, it does have flexion and
skewness.  Circles of constant radius from the earth are bent into
straight lines for example, and a rocket going out from the earth at
constant speed would be slowing down on the map.

In Figure~\ref{fg:us_mercator}, note that the continental United
States also appears lopsided on the Mercator map.  The geographical
center of the continental United States (which is in Kansas) appears
in the lower half of the continental United States on the Mercator map
because the scale factor on the map gets larger and larger the further
north one goes on the Map.  Thus, the continental United States is
lopsided toward the north in the Mercator map.  Flexion or bending is
manifest on the map as a bending of geodesics on the map, and skewness
or lopsidedness is manifest on the map as the midpoint of a geodesic
line segment on the map not being at the midpoint of that geodesic arc
as shown on the map.

\section{Goldberg-Gott Indicatrices}

\label{sec:gg}

We began this discussion with the virtues of the Tissot indicatrices.
Likewise, we have produced a simple indicatrix to show the flexion and
skewness in a map as well as the isotropy and area properties
indicated by the Tissot indicatrices.  We will refer to them as
``Goldberg-Gott" indicatrices.  These are constructed as follows.  At
a specific point on the map draw a circle on the globe of radius
$12^\circ$, and then plot it on the map.  Inside this circle, plot the
north-south, and east-west geodesics through the central point on the
map.  This leaves a $\oplus$ symbol on the map.  If the map were
perfect, this would be a perfect circle and the cross arms would be
perfectly straight, with their intersection at the center of the
circle.

\begin{figure}[h]
\centerline{\psfig{figure=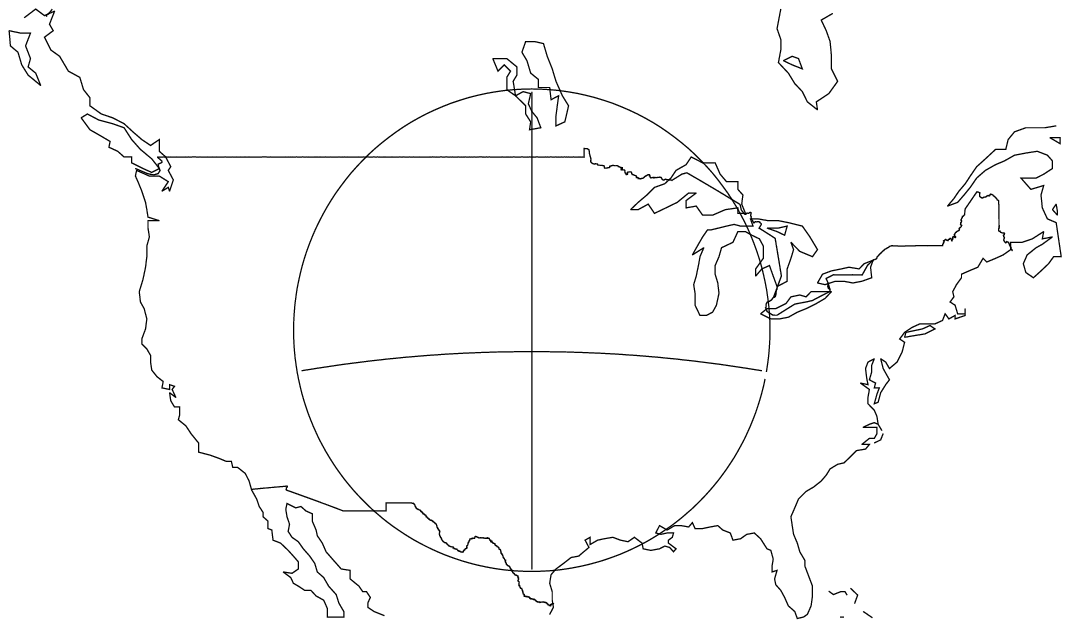,angle=0,height=3in}}
\caption{A Mercator projection cutout of continental the United
  States.  We have have put a Goldberg-Gott indicatrix (a circle of
  radius $12^\circ$ with N-S and E-W geodesics from the central point)
  at the geographic center of the continental U.S.}
\label{fg:us_mercator}
\vspace{0.2in}
\centerline{\psfig{figure=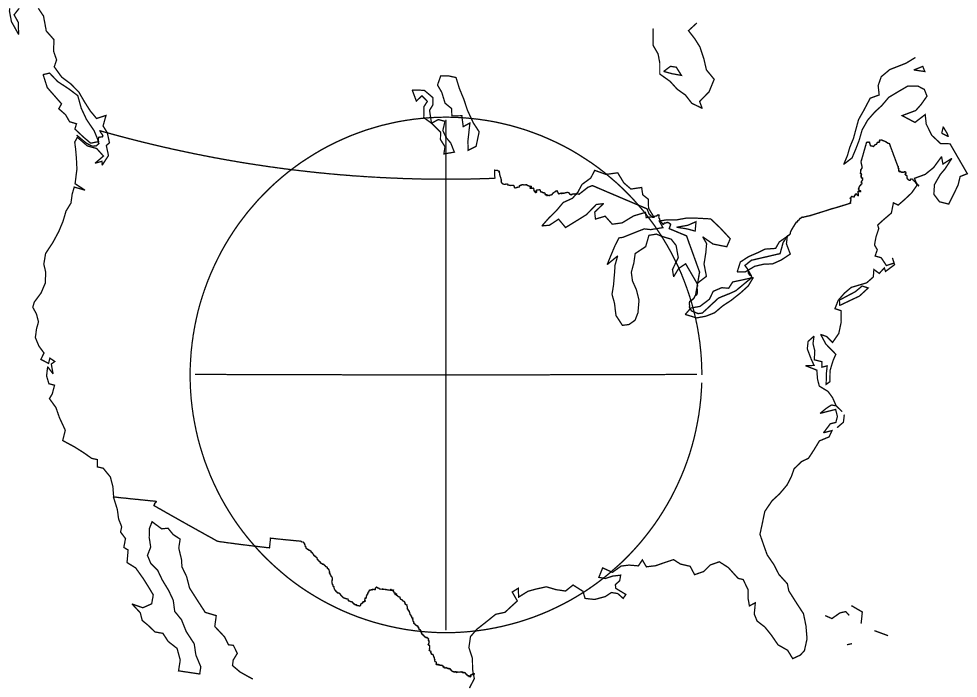,angle=0,height=3in}}
\caption{A oblique Mercator projection cutout of the United States.
  We have have put a Goldberg-Gott indicatrix at the geographic center
  of the continental U.S.  Notice that this projection has no flexion
  or skewness at the center.}
\label{fg:us_oblique_mercator}
\end{figure}

We have produced such a Goldberg-Gott indicatrix located at the
geographic center of the continental United States for using a
Mercator projection in Figure~\ref{fg:us_mercator}.

We note that our technique is similar to that of Tobler (1964),
who projected circles of 15 degrees in radius on to maps.  However, it
differs in that Tobler's circles do not have their centers marked nor
perpendicular geodesics drawn from their centers like our
indicatrices, and so do not convey the information on flexion and
skewness.  But they do show the finite shape distortions of the
circles themselves which our indicatrices also cover. Tobler includes
tables showing for particular locations on the sinusoidal projection,
the maximum and minimum scale radius and maximum difference of radial
directions for circles of various radii on the sphere (which address
the Tissot issues of size and isotropy on finite circles but not
flexion and skewness).  Tobler also calculated errors in miles and
angular errors for 300 random triangles within spherical circles of
various radii in different map projections, and considered area errors
for such random triangles within land areas in various world map
projections.

Using our indicatrices, one can see in Fig.~\ref{fg:us_mercator} that
the north-south geodesic is straight, but that the east-west geodesic
is bent downward.  This shows dramatically the flexion in this region
of the Mercator map.  One can even read off the average value of the
flexion by hand.  Take a protractor and measure the tangent to the
east-west geodesic at the two ends of the cross bar.  Measure the
difference in the angle orientation of the two.  That gives the
integrated flexion along $24^\circ$ of the globe.  Divide that angle
difference by $24^\circ$ and you will have the average value of the
flexion along that curve.

The skewness is also visible in that the center of the cross is below
the center of the circle, showing the lopsidedness to the north. In
fact, one can observe the skewness in any direction from the center by
seeing how far off center the center of the cross is with respect to
the center of the circle in different directions.  For comparison, we
have in Figure~\ref{fg:us_oblique_mercator} shown the continental
United States in a oblique Mercator projection where the east-west
geodesic through the geographic center of the continental United
States is now the equator of the Mercator projection. 

The flexion and skewness along the equator of a Mercator map are
indeed zero, so the arms of the cross are now straight, and the circle
is now nearly a perfect circle centered on the center of the cross.
This gives a "straight on" view of the continental United States, that
more accurately portrays its appearance on the globe. 

\begin{figure}[h]
\centerline{\psfig{figure=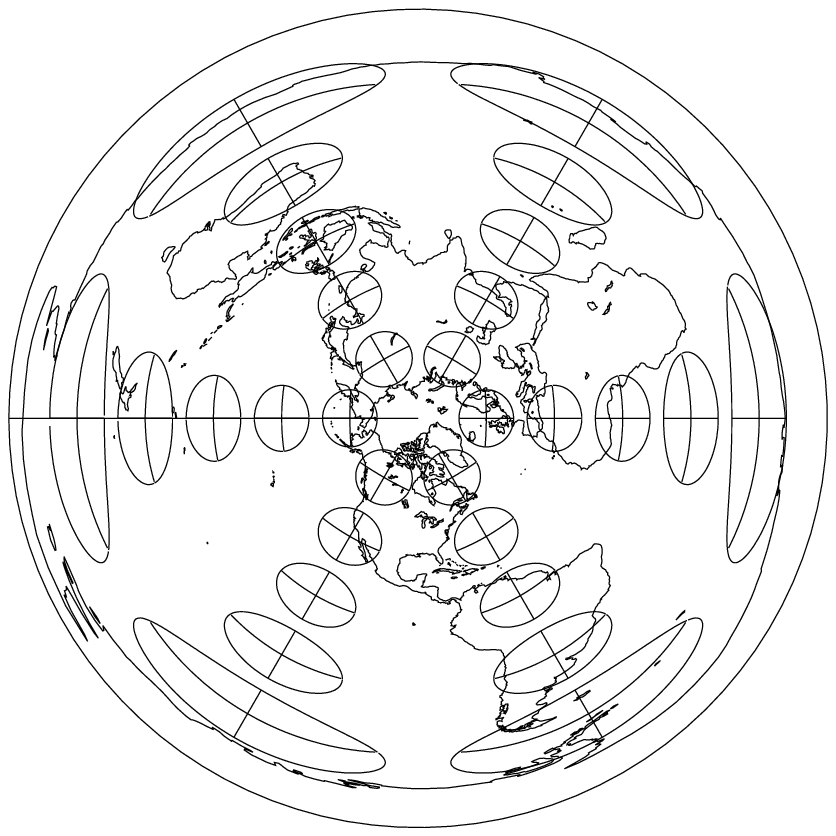,angle=0,height=3.7in}}
\caption{The Indicatrix map for an azimuthal equidistant projection.}
\label{fg:indmap_first}
\vspace{0.2in}
\centerline{\psfig{figure=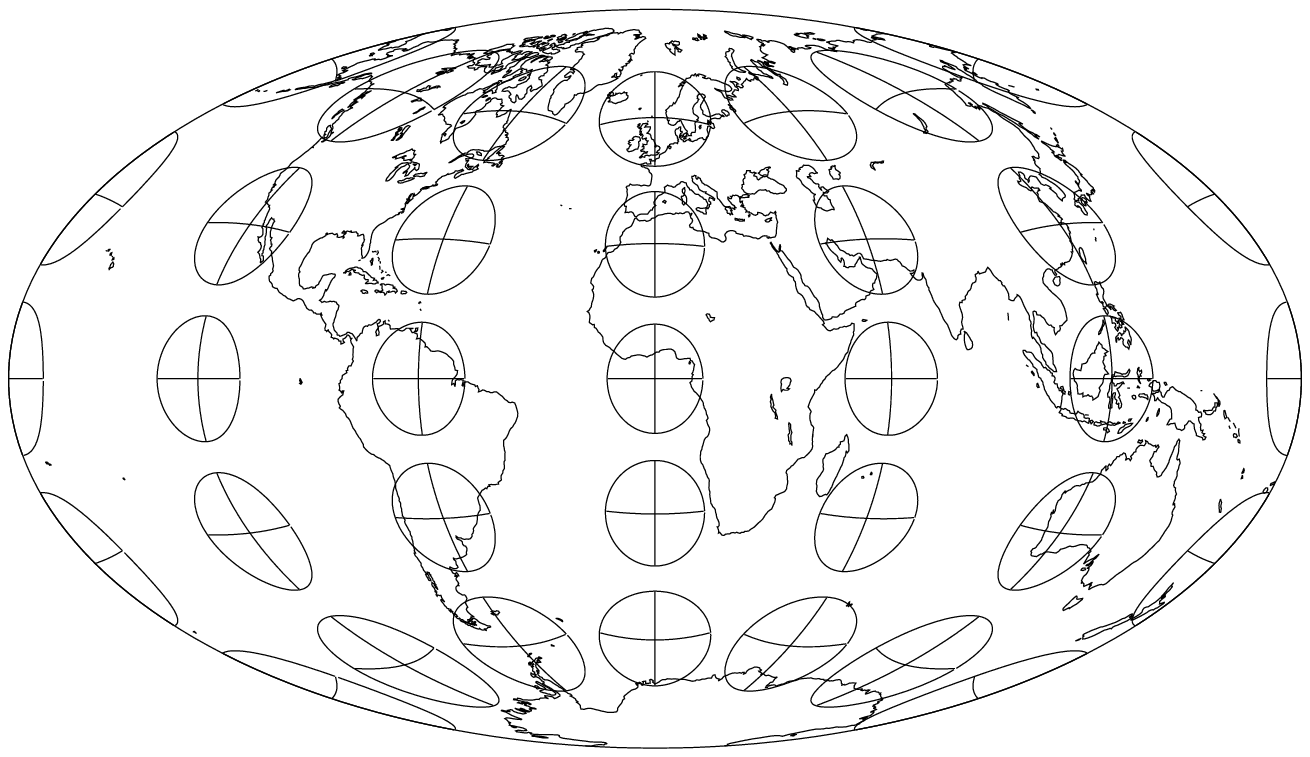,angle=0,height=3.7in}}
\caption{The Indicatrix map for a Briesemeister projection.}
\end{figure}

\begin{figure}[h]
\centerline{\psfig{figure=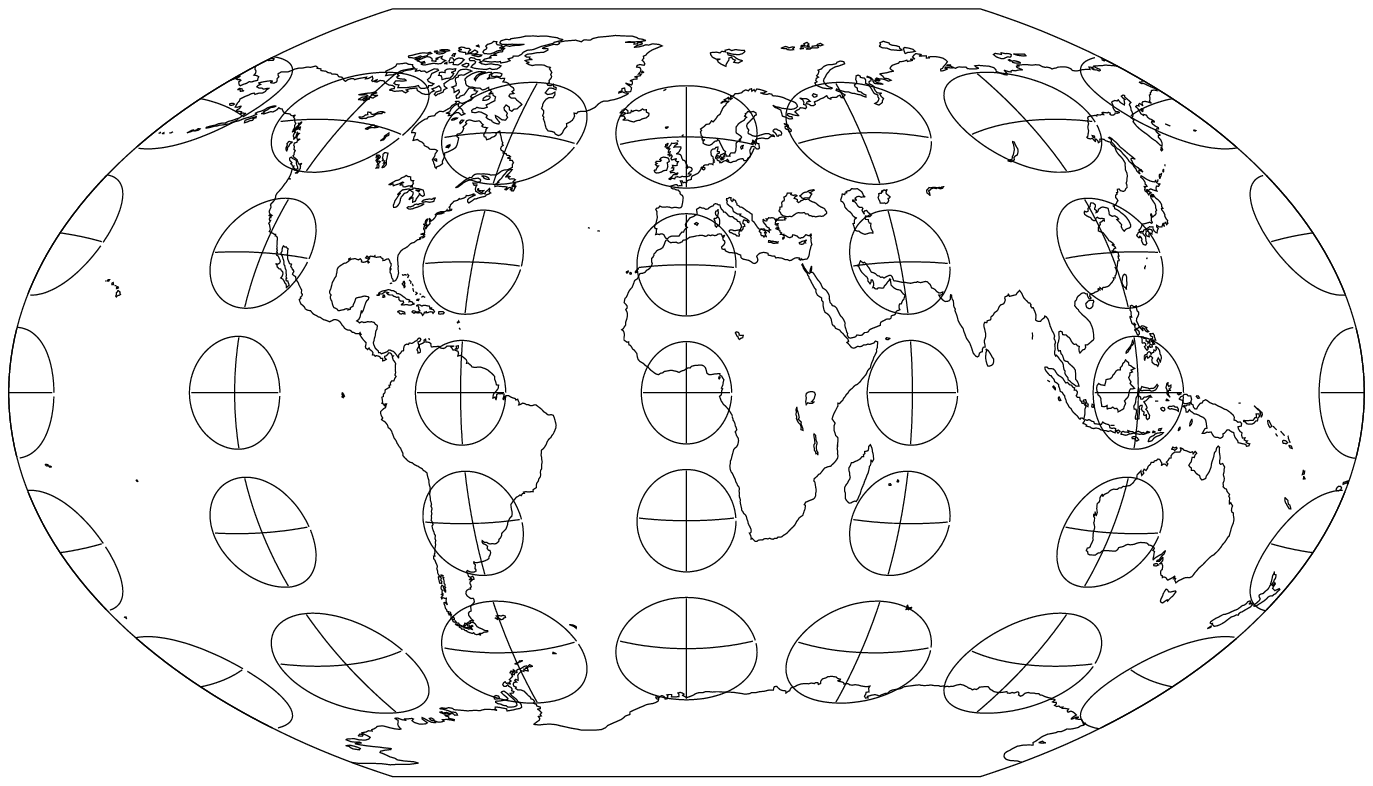,angle=0,height=3.7in}}
\caption{The Indicatrix map for a Winkel-Tripel (Times Atlas)
projection.}
\label{fg:indmap_last}
\end{figure}

One can place the Goldberg-Gott indicatrices every $60^\circ$ in
longitude and every $30^\circ$ in latitude on the globe to show how
the flexion and skewness vary over the map.  In
Figs.~\ref{fg:indmap_first}-~\ref{fg:indmap_last}, we provide G-G
indicatrix maps for a number of well-known projections.

In fact, the Goldberg-Gott almost indicatrices can just replace the
Tissot indicatrices because the shape and size of the oval in the
Goldberg-Gott indicatrix is approximately the size and shape of the
Tissot ellipse.  The Tissot ellipse shows the [magnified] shape and
size of an infinitesimal circle on the globe, the oval in the
Goldberg-Gott indicatrix $\oplus$ shows the shape and size of a finite
circle (radius $12^\circ$) on the map itself at correct scale.  Thus,
if the map is equal area, the Goldberg-Gott indicatrices, will all
have equal area on the map.  If the map is conformal the Goldberg-Gott
indicatrices will all be nearly perfectly circular.  If there is a 2:1
anisotropy in the Tissot ellipses in a given region the Goldberg-Gott
indicatrices ovals will have that same 2:1 axis ratio.

\section{A differential geometry approach}

\label{sec:geometry}

Thus far, we have defined the general properties of skewness and
flexion, given a few analytic results for particular map projections,
and given a graphic approach for describing and interpreting flexion
and skewness on maps.  In this section, we approach the matter
somewhat differently, and produce general analytic results for all
projections as well as a prescription for measuring the flexion and
skewness analytically.  

\subsection{Coordinate Transforms}

Let's consider a spherical globe with coordinates:
\begin{equation}
x^{\ol{a}}=\left(
\begin{array}{c}
\phi\\
\lambda
\end{array}\right)\ ,
\end{equation}
Note that here and throughout, we will use $x^{\ol{a}}$ to refer to
coordinates in the globe frame, and $x^a$ to refer to coordinates in
the map frame.

On the globe, the metric is:
\begin{equation}
g_{\ol{a}\ol{b}}=\left(
\begin{array}{cc}
1&0\\
0&\cos^2\phi\\
\end{array}
\right)
\end{equation}
such that, as always, the distance between two points can be expressed
as:
\begin{equation}
dl^2=dx^{\ol{a}}dx^{\ol{b}}g_{\ol{a}\ol{b}}
\end{equation}

Now, consider an arbitrary 2-d coordinate transformation:
\begin{equation}
\left( 
\begin{array}{c}
x \\
y\\ 
\end{array}
\right)=\left(
\begin{array}{c}
x^1 (\phi,\lambda)\\
x^2 (\phi, \lambda)
\end{array}
\right)
\end{equation}
Of course, from this definition, we may easily compute a local
transformation matrix:
\begin{equation}
\Lambda^a_{\ol{a}}=\frac{\partial x^a}{\partial{x^{\ol{a}}}}
\label{eq:transform}
\end{equation}

The inverse matrix is $\Lambda^{\ol{a}}_a=\partial x^{\ol{a}}/\partial
x^a$. The metric in the map frame is:
\begin{equation}
g_{ab}= \Lambda^{\ol{a}}_a \Lambda^{\ol{b}}_b g_{\ol{a}\ol{b}}
\end{equation}
From this, we may then compute
the Christoffel Symbols of general relativity:
\begin{equation}
\Gamma^{a}_{bc}=\frac{1}{2} g^{ae}\left(
g_{eb,c}+g_{ec,b}-g_{bc,e}
\right)
\end{equation}
where standard convention tells us to sum over identical indices in
the upper and lower positions, and where a comma indicates a partial
derivative with respect to a coordinate.

In practice, actually computing the Christoffel symbols for an
arbitrary projection is not simple.  To do this analytically requires
that we have an analytic form for the map inversion.  However, we make
available a numerical code to compute the Christoffel symbols for all map
projections discussed in this work on our projections website (see
below).

\subsection{Analytic forms of Flexion and Skewness}

The whole point of computing the Christoffel symbols is that we want
to address a very simple question: How are large structures distorted
when projected onto a map?  Clearly to an observer on the globe, a
straight line is easy to generate.  Point in a particular direction,
and start driving (assuming your car can drive on the ocean) with the
steering wheel set straight ahead.  Drive for a fixed distance in
units of angles or radian.  Record all points along the way.

Geometers, of course, know this route as a geodesic, and if we
consider $\tau$ to represent a physical distance on the surface of
the earth, then the geodesic equation may be expressed as:
\begin{equation}
\frac{du^a}{d\tau}=-\Gamma^{a}_{bc}u^b u^c
\label{eq:geodesic}
\end{equation}
where 
\begin{equation}
u^a=\frac{dx^a}{d\tau}\ .
\end{equation}

Equation~(\ref{eq:geodesic}) describes the bending
of straight lines on a particular map projection, and thus, if all of
the Christoffel symbols could vanish, we would clearly have a perfect
Cartesian map.  Not possible, of course.

But what is the physical significance of the Christoffel symbols?
Since the lower indices are symmetric by inspection, there are 6
unique symbols.  What do they mean?

\subsubsection{Analytic Flexion}

We define a vector oriented in a particular direction:
\begin{equation}
\tilde{u}^a(\theta)=\left(
\begin{array}{c}
\cos\theta\\
\sin\theta
\end{array}
\right)
\end{equation}
In reality, this is {\it not} the unit vector, since:
\begin{equation}
|\tilde{u}|^2=g_{ab}\tilde{u}^a \tilde{u}^b \neq 1
\end{equation}
Of course, we {\it could} define a true unit unit vector:
\begin{equation}
u(\theta)=l(\theta)\tilde{u}(\theta)
\end{equation}
where $l$ is the ``length'' in grid coordinates of the unit vector.
This has a value of:
\begin{equation}
l(\theta)=\frac{1}{\sqrt{\cos^2\theta g_{11}+\sin^2\theta
    g_{22}+2\sin\theta\cos\theta g_{12}}}
\end{equation}
Of course, it is clear that at any point on the map, the set of all
$u(\theta)$ represents an ellipse -- the Tissot ellipse.  

We define the flexion in the following manner: Follow a particular
geodesic a distance, $d\tau$ (measured in, for example, radians).  On
the map, the geodesic will change direction by an angle $d\theta$.  The ratio:
\begin{equation}
f=\frac{d\theta}{d\tau}
\end{equation}
is the flexion.  Note that for all polar projections, the equator will
have a flexion of 1.  

In terms of the geodesic equation, the flexion can be expressed as:
\begin{eqnarray}
f(\theta)&=&l(\theta)\left(\frac{d{\bf \tilde{u}}}{d\tau} \times
\tilde{{\bf u}} \right)\nonumber\\
&=&l(\theta)\left(\Gamma_{ab}^2 \tilde{u}^a \tilde{u}^b \tilde{u}^1 -
\Gamma_{ab}^1 \tilde{u}^a \tilde{u}^b \tilde{u}^2\right)\nonumber \\
\label{eq:flexion}
\end{eqnarray}
where ${\bf \tilde{u}}$ denotes a vector. 

Some light can be shed on equation~(\ref{eq:flexion}) if we expand the
expression explicitly into trigonometric functions:
\begin{eqnarray}
f(\theta)&=&l(\theta)\left[\Gamma^{2}_{11}\cos^3\theta+(2\Gamma^{2}_{12}-\Gamma^{1}_{11})\cos^2\theta\sin\theta\right.
  \nonumber\\ 
&&\left.-(2\Gamma^{1}_{12}-\Gamma^{2}_{22})\sin^2\theta\cos\theta-\Gamma^{1}_{22}\sin^3\theta\right]
\label{eq:flexion2}
\end{eqnarray}
This immediately reveals an important idea.  If:
\begin{eqnarray}
\Gamma^{2}_{11}&=&\Gamma^{1}_{22}=0\nonumber \\
\Gamma^{1}_{11}&=&2\Gamma^{2}_{12}\nonumber \\
\Gamma^{2}_{22}&=&2\Gamma^{1}_{12}
\label{eq:straight}
\end{eqnarray}
then $f\theta)$ in all directions will be zero.  That is, all
geodesics will be straight lines in this projection.  As we will see,
this is true for the Gnomonic projection.

Inspection of equation~(\ref{eq:flexion2}) shows it is clearly
anti-symmetric about exchanges of $\theta$ and $-\theta$, since
$\theta$ is {\it not} the normal vector to the geodesic, but rather
runs along it.

\subsubsection{Analytic Skewness}

Essentially, a skewness means that if you walk (initially) north (for
example) for 1000 miles, or walk south for 1000 miles, you will cover
different amounts of map coordinate.

The skewness along a geodesic can be defined similarly to the flexion:
\begin{eqnarray}
s(\theta)&=&l(\theta)\left(\frac{d{\bf \tilde u}}{d\tau}\cdot {\bf
  \tilde{u}}\right)\nonumber \\
&=&l(\theta)\left(\Gamma^1_{bc}\tilde{u}^b\tilde{u}^c
\tilde{u}^1+\Gamma^2_{bc}\tilde{u}^b \tilde{u}^c\tilde{u}^2\right)
\end{eqnarray}
As with the flexion, we may expand these out explicitly:
\begin{eqnarray}
s(\theta)&=&l(\theta)\left[ \Gamma^{1}_{11} \cos^3\theta +
(2\Gamma^{1}_{12}+\Gamma^{2}_{11})\cos^2\theta\sin\theta \right.
    \nonumber \\
&& \left. +(2\Gamma^{2}_{12}+\Gamma{1}_{22})\sin\theta
\cos^2\theta+\Gamma^{2}_{22}\sin^3\theta\right]
\end{eqnarray}
A map with no skewness will have:
\begin{eqnarray}
\Gamma^{1}_{11}&=&\Gamma^{2}_{22}=0\nonumber \\
\Gamma^{2}_{11}&=&-2\Gamma^{1}_{12}\nonumber \\
\Gamma^{1}_{22}&=&-2\Gamma^{2}_{12}
\label{eq:unskewed}
\end{eqnarray}

Unlike the flexion, we know of no projections with zero skewness everywhere.

\subsection{Projections with straightforward analytic results}

\subsubsection{The Gnomonic Projection}

The Gnomonic Projection is particularly interesting.  It has a
coordinate transformation:
\begin{equation}
\left(
\begin{array}{c}
x\\
y
\end{array}
\right)_{g}=
\left(
\begin{array}{c}
\cot\phi\cos\lambda\\
\cot\phi \sin\lambda
\end{array}
\right)
\end{equation}
which is directly invertible to yield:
\begin{equation}
\left(
\begin{array}{c}
\phi \\
\lambda
\end{array}
\right)=
\left(
\begin{array}{c}
\cos^{-1}\sqrt{\frac{x^2+y^2}{1+x^2+y^2}}\\
\tan^{-1}\left(\frac{y}{x}\right)
\end{array}
\right)
\end{equation}

The coordinate transformation is thus:
\begin{equation}
\Lambda^{a}_{\ol{a}}=
\left(
\begin{array}{cc}
-\frac{x}{\sqrt{x^2+y^2}(1+x^2+y^2)} &
 -\frac{y}{\sqrt{x^2+y^2}(1+x^2+y^2)} \\
-\frac{y}{x^2+y^2} & \frac{x}{x^2+y^2}
\end{array}
\right)
\end{equation}

We thus have the map metric:
\begin{equation}
g_{ab}=\left(
\begin{array}{cc}
\frac{1+y^2}{(1+x^2+y^2)^2} & \frac{-xy}{(1+x^2+y^2)^2} \\
 \frac{-xy}{(1+x^2+y^2)^2} & \frac{1+x^2}{(1+x^2+y^2)^2}
\end{array}
\right)
\end{equation}

We can compute the Christoffel symbols in the normal way. We find:
\begin{eqnarray}
\Gamma^{1}_{11}&=&-\frac{2x}{1+x^2+y^2}\nonumber \\
\Gamma^{2}_{11}&=&0\nonumber \\
\Gamma^{1}_{12}&=&-\frac{y}{1+x^2+y^2}\nonumber \\
\Gamma^{2}_{12}&=&-\frac{x}{1+x^2+y^2}\nonumber \\
\Gamma^{1}_{22}&=&0\nonumber \nonumber \\
\Gamma^{2}_{22}&=&-\frac{2y}{1+x^2+y^2}
\end{eqnarray}
This clearly satisfies the requirements of
equations~(\ref{eq:straight}), but not (\ref{eq:unskewed}), and thus,
the Gnomonic projection produces straight, but skewed geodesics.

\subsubsection{Stereographic}

The Stereographic projection is conformal, and thus, all of the Tissot
ellipses are circles.  Does this mean there is no skewness in the
projection?  No, as we've already seen.  The Stereographic projection
has the coordinate transformation:
\begin{equation}
\left(
\begin{array}{c}
x\\
y
\end{array}
\right)_{g}=
\left(
\begin{array}{c}
\tan(\pi/4+\phi/2)\cos\lambda\\
\tan(\pi/4+\phi/2)\sin\lambda
\end{array}
\right)
\end{equation}
which, again, is directly invertible to yield:
\begin{equation}
\left(
\begin{array}{c}
\phi \\
\lambda
\end{array}
\right)=
\left(
\begin{array}{c}
\frac{\pi}{2}-2\tan^{-1}(\sqrt{x^2+y^2})\\
\tan^{-1}\left(\frac{y}{x}\right)
\end{array}
\right)
\end{equation}

The coordinate transformation is:
\begin{equation}
\Lambda^{a}_{\ol{a}}=
\left(
\begin{array}{cc}
-\frac{2x}{\sqrt{x^2+y^2}(1+x^2+y^2)} &
 -\frac{2y}{\sqrt{x^2+y^2}(1+x^2+y^2)} \\
-\frac{y}{x^2+y^2} & \frac{x}{x^2+y^2}
\end{array}
\right)
\end{equation}

This can be used to compute the metric on the map:
\begin{equation}
g_{ab}=\left(
\begin{array}{cc}
\frac{4}{(1+x^2+y^2)^2} & 0 \\
0& \frac{4}{(1+x^2+y^2)^2}
\end{array}
\right)
\end{equation}
This clearly indicates that all Tissot ellipses will be circular.  

From these, of course, we can compute the Christoffel symbols:
\begin{eqnarray}
\Gamma^{1}_{22}&=&-\Gamma^{1}_{11}=-\Gamma^{2}_{12}=\frac{2x}{1+x^2+y^2}\\
\Gamma^{2}_{11}&=&-\Gamma^{2}_{22}=-\Gamma^{1}_{12}=\frac{2y}{1+x^2+y^2}
\end{eqnarray}
It is clear from inspection that geodesics are generally neither
straight nor unskewed.  

Moreover, it is clear that $l(\theta)$ is independent of orientation
since the map is conformal.  In general it can be shown to be:
\begin{equation}
l=\frac{1}{2}(1+x^2+y^2)\ .
\end{equation}
In the stereographic projection, a circle of radius $12^\circ$ on the
globe is a  perfect circle on the map but the center of the circle on
the globe is not at the center of the circle on the globe (see
Figure~\ref{fg:stereographic}), and thus, there is skewness.

\section{Discussion}

\label{sec:discussion}

\subsection{Numerical Analysis of Standard Map Projections}

\label{sec:numerical}

Not all projections produce such simple results.  Thus, in general, we
will want to compute the local flexion and skewness numerically.  Our
approach is as follows: For each projection we chose 30,000 points
selected randomly on the surface of a globe.  For each of these
points, we chose a random direction to start a geodesic.  We follow
that geodesic using small steps ($d\tau\simeq 10^{-5}$ rad)
numerically, and use standard difference methods to compute the
map velocity and acceleration along the geodesic.  We are thus able to
compute the metric and the Christoffel symbols (and thus the flexion
and skewness) directly.  We make our IDL (Interactive Data Language)
code available to the interested reader at our projection webpage (see
below).  Likewise, we also do a distance test, in which pairs of
points, (i,j), are chosen at random and the distance is measured both
on the globe and on the map.

This is a somewhat different perspective than simply inspecting the
Goldberg-Gott indicatrices at a few locations, since we are now doing
a uniform sample over the surface of the globe, rather than a uniform
sampling over the {\it map}.  When looking at the indicatrix map we
can occasionally get a distorted view as to the quality of a
particular projection.   Some (like the Mercator) have relatively good
fits over most of the globe, but the high latitudes can, in principle,
be projected to infinite areas, and thus, the reader may erroneously
think the Mercator infinitely bad.  By sampling uniformly over the
globe, we get a fair assessment of the overall quality of a
particular projection.

We define a number of fit parameters: I, corresponding to errors in
the local isotropy (zero for conformal projections), A, corresponding
to errors in the Area (zero for equal area projections), F,
corresponding to flexion (defined in the discussion of flexion,
above), S, corresponding to skewness (also defined above), D,
corresponding to distance errors, and B, corresponding to the average
number of map boundary cuts crossed by the shortest geodesic
connecting a random pair of points.
\begin{eqnarray}
I&=&RMS\left(\ln \frac{a_i}{b_i}\right)\\
A&=& RMS\left( \ln a_i b_i-\langle \ln a_i b_i \rangle  \right)\\
F&=& \langle |f_i|\rangle\\
S&=& \langle |s_i|\rangle\\
D&=& RMS\left( \ln\frac{d_{ij,map}}{d_{ij,globe}} \right)\\
B&=&\frac{L_{B}}{4\pi}
\end{eqnarray}
where $a_i$ and $b_i$ are the major and minor axes of the local Tissot
ellipses of random point, $i$, $\langle X_i \rangle$ indicates the
mean of property, X, and $L_B$ is the total length of the boundary cuts.  

Logarithmic errors have been used before by Kavrayskiy (1958).
Bugayevskiy and Snyder (1995) adopted schemes that weighted
isotropy errors, $(a/b -1)^2$, and area errors, $(ab - 1)^2$,
according to one's interest in isotropy and area. Airy (1861) had just
given these terms equal weight.  

\begin{table}[h!]
\begin{tabular}{|l|c c c c c c|}
\hline 
Projection & I & A & F&  S & D  & B\\ \hline\hline
{\bf Non Interrupted Projections} & & & & & &\\ 
Azimuthal Equidistant & 0.87 & 0.60 & 1.0 & 0.57 & 0.356 & 0 \\
Gott-Mugnolo Azimuthal          & 1.2  & 0.20 & 1.0 & 0.59 & 0.341 & 0\\
Lambert Azimuthal     & 1.4  & 0    & 1.0 & 2.1  & 0.343 & 0 \\
Stereographic         & 0    & 2.0  & 1.0 & 1.0  & 0.714 & 0\\ \hline
{\bf 1 180 deg. Boundary Cut} & & & & & & \\
Breisemeister   & 0.79 & 0   & 0.81 & 0.42 & 0.372  & 0.25\\
Eckert IV       & 0.70 & 0   & 0.75 & 0.55 & 0.390 & 0.25\\
Eckert VI       & 0.73 & 0   & 0.82 & 0.61 & 0.385 & 0.25 \\
Equirectangular & 0.51 & 0.41& 0.64 & 0.60 & 0.449 & 0.25 \\
Gall-Peters     & 0.82 & 0   & 0.76 & 0.69 & 0.390 & 0.25 \\
Gall Stereographic & 0.28 & 0.54 & 0.67 & 0.52 & 0.420 & 0.25\\
Gott Elliptical & 0.86 & 0   & 0.85 & 0.44 & 0.365 & 0.25 \\
Gott-Mugnolo Elliptical & 0.90 & 0 & 0.82 & 0.43 & 0.348 & 0.25\\
Hammer          & 0.81 & 0   & 0.82 & 0.46 & 0.388 & 0.25 \\ 
Hammer-Wagner   & 0.687 & 0 & 0.789 & 0.518 & 0.377 & 0.25 \\
Kavrayskiy VII  & 0.45 & 0.31& 0.69 & 0.41 & 0.405 & 0.25 \\
Lagrange        & 0    & 0.73& 0.53 & 0.53 & 0.432 & 0.25 \\
Lambert Conic   & 0    & 1.0 & 0.67 & 0.67 & 0.460 & 0.25 \\
Mercator        & 0    & 0.84& 0.64 & 0.64 & 0.440 & 0.25 \\
Miller          & 0.25 & 0.61& 0.62 & 0.60 & 0.439 & 0.25 \\
Mollweide       & 0.76 & 0   & 0.81 & 0.54 & 0.390 & 0.25 \\
Polyconic       & 0.79 & 0.49& 0.92 & 0.44 & 0.364 & 0.25 \\
Sinusoidal      & 0.94 & 0   & 0.84 & 0.68 & 0.407 & 0.25 \\
Winkel-Tripel   & 0.49 & 0.22& 0.74 & 0.34 & 0.374 & 0.25 \\ 
Winkel-Tripel (Times) & 0.48 & 0.24 & 0.71 & 0.373 & 0.39 & 0.25\\
\hline

{\bf 1 360 deg. Boundary Cut} && & & & &\\
Lambert Azimuthal (2 hemisphere) & 0.36 & 0    & 0.52 & 0.11 & 0.432&0.5\\
Stereographic (2 hemisphere)     & 0    & 0.39 & 0.37 & 0.37 & 0.692 &0.5\\    
\hline

{\bf Multiple Boundary Cut Projections} & & & & & & \\

Gnomonic Cube   & 0.22 & 0.37 &   0.12  &   0.87  &  0.43 & 0.686 \\\hline
\end{tabular}
\label{tab:global}
\caption{The errors in isotropy, area, flexion, skewness, distances
 , and boundary cuts for some standard projections.}
\end{table}

In Table~\ref{tab:global} we compare a these measures for a number of
standard projections, which, for fairness of comparison we divide into
a number of categories.  The Gott-Elliptical, The Gott-Mugnolo
Elliptical, and the Gott-Mugnolo Azimuthal have been discussed in
Gott, Mugnolo \& Colley (2007) and Gott, et al. (2007), where they
have been applied to the earth, Mars, the moon, and the Cosmic
Microwave all sky map.  

First, we show projections which represent the
complete globe without interrupts.  These projections are azimuthal
and the average flexion over these maps is 1.

Second, we show the set of whole earth projections with one
$180^\circ$ interrupt.  These include rectangular and elliptical
projections.  Note that among all of the complete projections with 0
or 1 $180^\circ$ interrupt, there are a number of ``winners'' with
regards to performance for flexion and skewness.  The Lagrange has the
smallest flexion.  The Winkel-Tripel has the smallest skewness.  For
all conformal projections, the skewness is equal to the flexion.

Of all of the whole earth projections, the most accurate for distance
measure between points is the Gott-Mugnolo Azimuthal, followed very closely by
the Lambert Azimuthal (Gott, Mugnolo \& Colley 2007).

In the third and fourth groups, we show 2-hemisphere and other
multiple cut projections, respectively.

In the final group, we have a projection with multiple interrupts, the
gnomonic cube, which is defined piecemeal.  This is a particularly
interesting projection since the gnomonic is locally flexion-free, but
it is clear that geodesics will {\it not} trace out straight lines in
the gnomonic cube cube map because they bend when they cross an edge
between faces.  The gnomonic cube is presented as a cross, so 5 edges
are included in the map proper.  Geodesics bend when they cross an
edge in this laid out cross configuration.

The above comparisons do not depend on how important each of the
criteria are (i.e. what weighting to give each measure).  Laskowski (1997ab)
suggested a means of ranking very disparate maps.
Though his weighting scheme is not unique, as a simple illustration of
how this can be done, we will minimize the sum of the squares of all
6 parameters, normalized to their values in the equirectangular projection:
\begin{equation}
\Sigma_\varepsilon=\left(\frac{I}{N_i}\right)^2+
\left(\frac{A}{N_a}\right)^2+ \left(\frac{F}{N_f}\right)^2+
\left(\frac{S}{N_s}\right)^2+ \left(\frac{D}{N_d}\right)^2+
\left(\frac{B}{N_b}\right)^2
\end{equation}
Following Laskowksi (1997ab) we set the normalization
constants equal to the values of these errors in the Equirectangular
projection ($x=\lambda$, $y=\phi$): $N_i= 0.51$, $N_a = 0.41$, $N_f =
0.64$, $N_s = 0.60$, $N_d = 0.449$, $N_b = 0.25$.

The projections with the lowest values of $\Sigma_\varepsilon$ are:
\begin{enumerate}
\item Winkel-Tripel       (4.5629)
\item Winkel-Tripel (Times Atlas)  (4.5687)
\item Kavrayskiy VII  (4.8390)
\item Gall Stereographic (5.7582)
\item Hammer-Wagner (5.7847)
\item Eckert IV       (5.8519)
\end{enumerate}
This approach is certainly not unique.  One may take issue with the
weighting of the individual parameters, the domain over which they are
applied (the whole earth, as opposed to continents only, for example),
or even how the parameters are computed, we present it as a simple
example of how our results may be combined with previous studies of
map projections.  It is interesting that the ``best'' map, as selected
by this criterion is the one already used by the National Geographic
for it's whole world projection.  It is also interesting to note that
the Winkel-Tripel has especially low skewness.

Another nice property of this weighting scheme is that it gives a low
score to pathological projections.  For example, a series of n gores
(made using the polyconic projection) arranged in a sunflower pattern
would approximate the Azimuthal equidistant projection in distance
errors as n became large but would have arbitrarily low values of I,
A, F, and S. But if a boundary cut term B is included this term would
blow up and save us from picking the bad subdivided map as better than
the more visually pleasing projections described above.  Stretching
individual pixels at the edges of the gores to fill the gaps between
them would eliminate the boundary cuts.  However, it would also cause
the skewness to blow up, preventing us from giving a good score to a
bad projection.

Interested readers may visit {\tt
www.physics.drexel.edu/$\sim$goldberg/projections/} to download a free
IDL code to measure the flexion, area, and other measures discussed in
this paper.  We have not done all known projections, but have covered
ones that have available mathematical formulas and we thought likely
to do well.

\subsection{Conclusions}

We have developed two new measures of curvature distortions found in
maps of the earth.  The Skewness and the Flexion can be used to
identify particularly warped sections of a map, or to identify
features (such as the U.S. Canada border), which appear as a straight
line on some map projections but do not, in fact, follow geodesics.
We have developed a new graphical tool, called the Goldberg-Gott
indicatrices, which can be used to show these distortions, along with
those in area and shape, at many points in a map, and have produced
indicatrix maps for a number of popular projections.  Finally, we have
used these measures to produce global distortion measures for
different projections.  We have found that the Winkel-Tripel produces
low distortion on most measures, and, in particular, has the lowest
skewness of all projections in our sample with a $180^\circ$ boundary cut.

\section*{Acknowledgments}

JRG is supported by National Science Foundation grant AST04-06713.
DMG is supported by a NASA Astrophysics Theory Grant.  We thank Wes
Colley for useful discussions, as well as the helpful comments made by
the anonymous referees.

\newpage

\newpage

\newpage
\begin{figure}[h]
\centerline{\psfig{figure=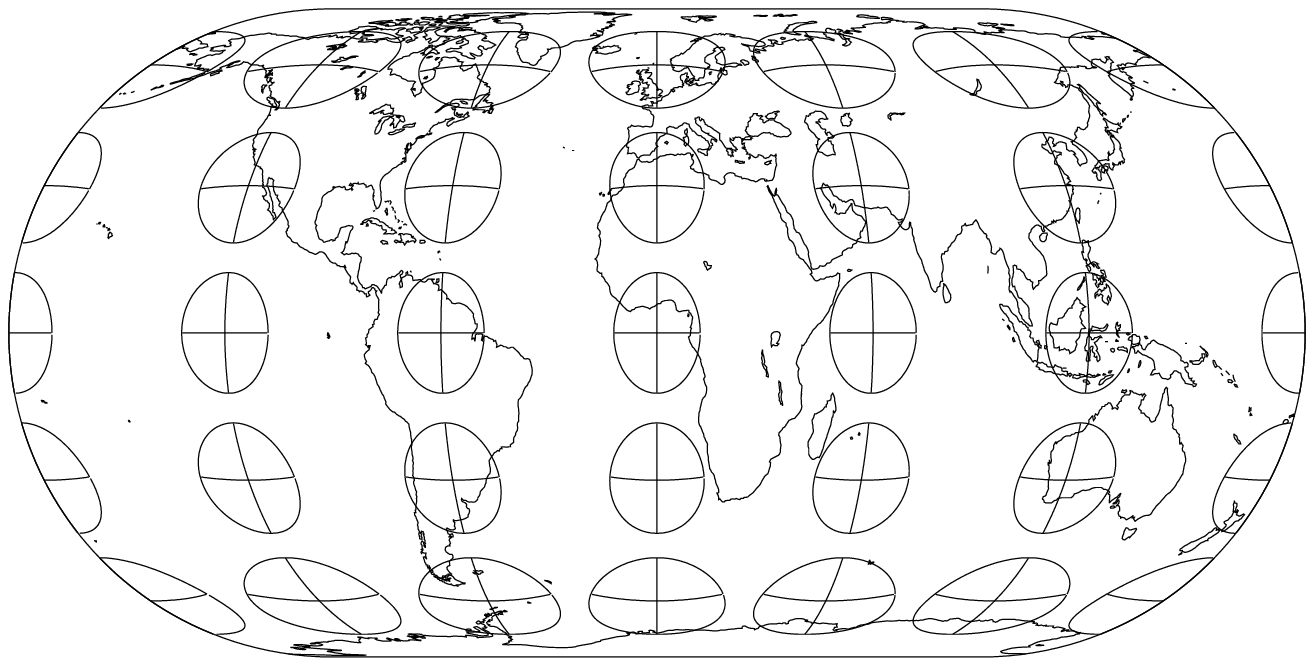,angle=0,height=3.7in}}
\caption{The Indicatrix map for an Eckert IV projection.}
\vspace{0.2in}
\centerline{\psfig{figure=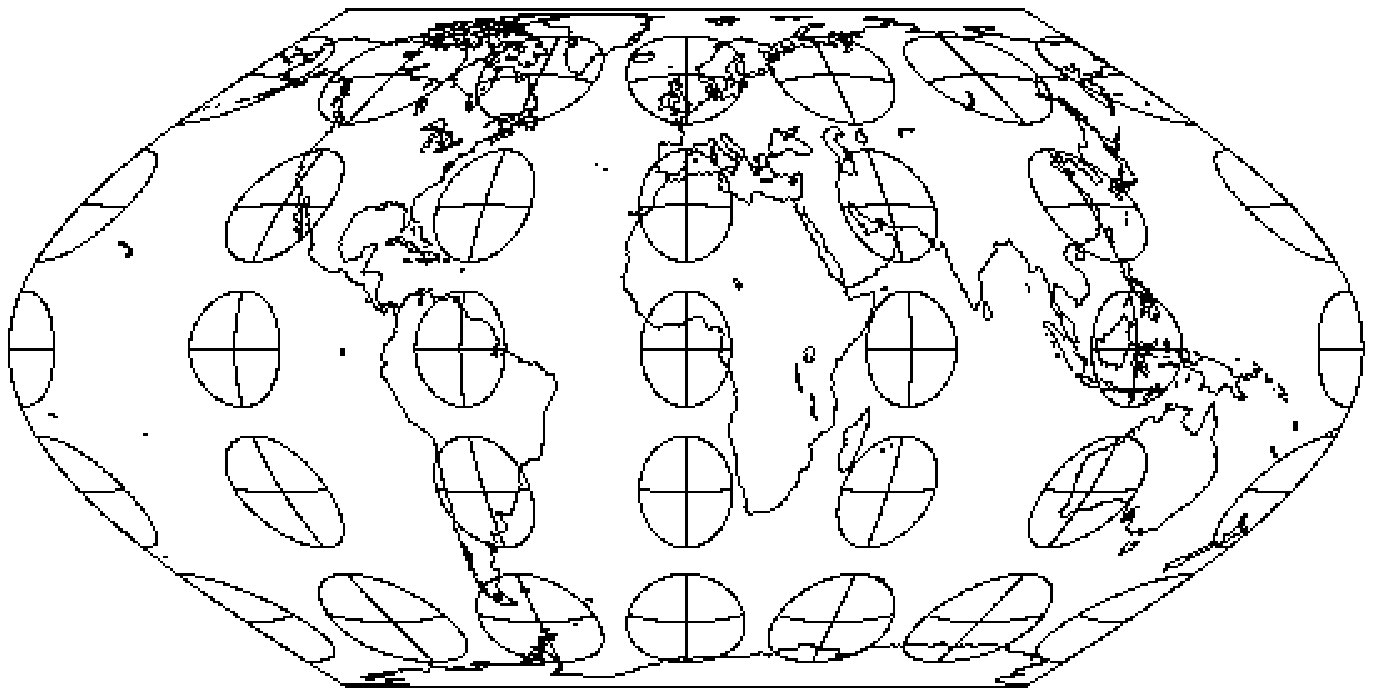,angle=0,height=3.7in}}
\caption{The Indicatrix map for an Eckert VI projection.}
\end{figure}

\newpage

\begin{figure}[h]
\centerline{\psfig{figure=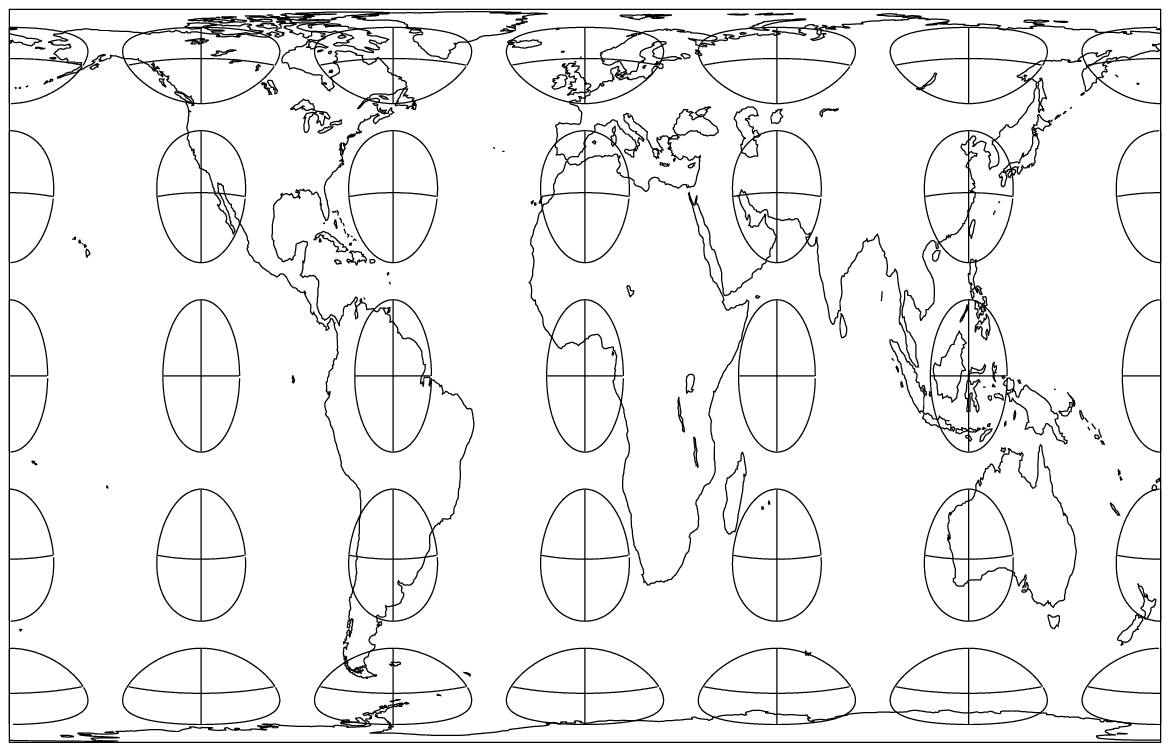,angle=0,height=3.7in}}
\caption{The Indicatrix map for a Gall-Peters projection.}
\vspace{0.2in}
\centerline{\psfig{figure=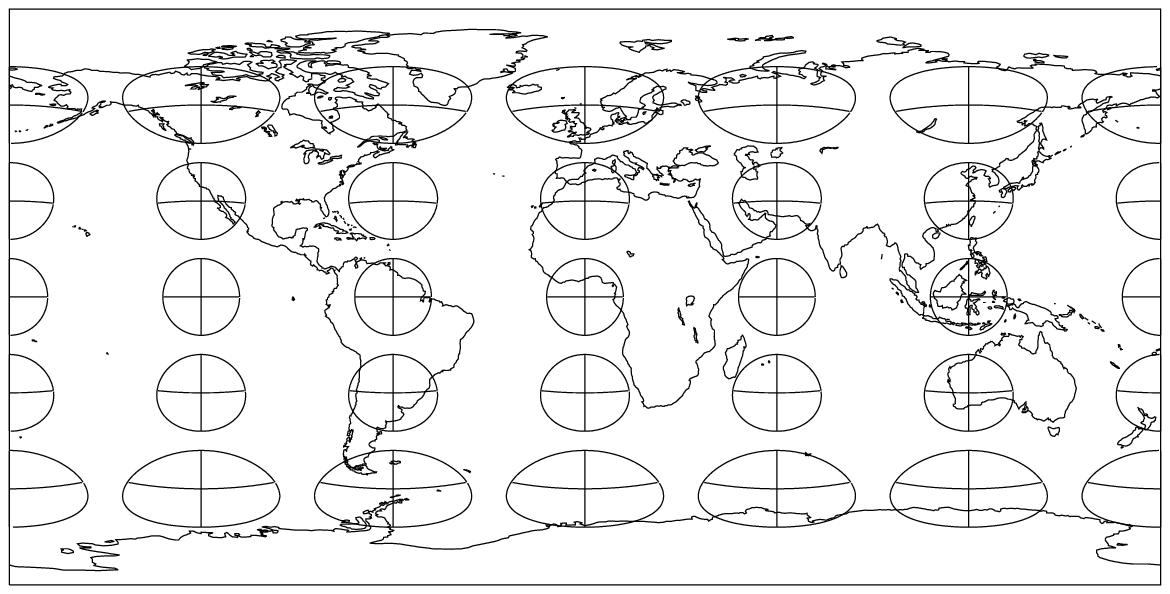,angle=0,height=3.7in}}
\caption{The Indicatrix map for an Equirectangular projection.}
\end{figure}

\newpage

\begin{figure}[h]
\centerline{\psfig{figure=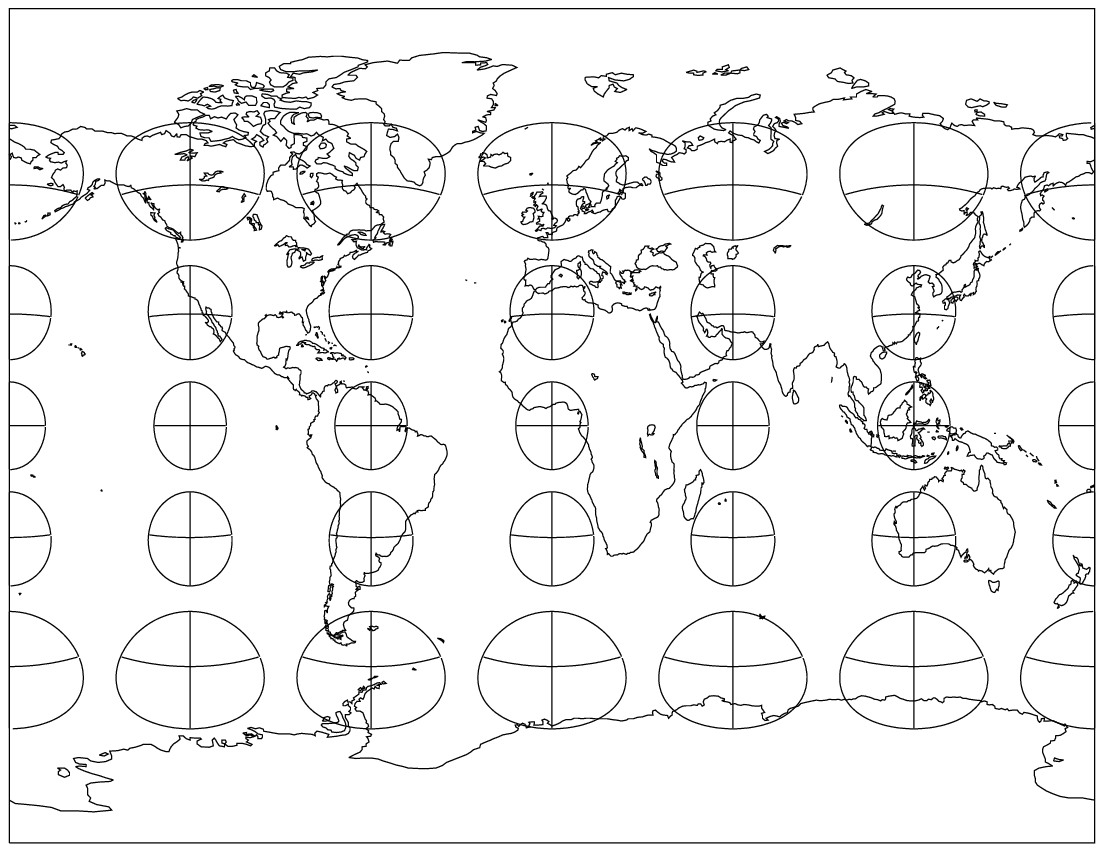,angle=0,height=3.7in}}
\caption{The Indicatrix map for a Gall Stereographic projection.}
\vspace{0.2in}
\centerline{\psfig{figure=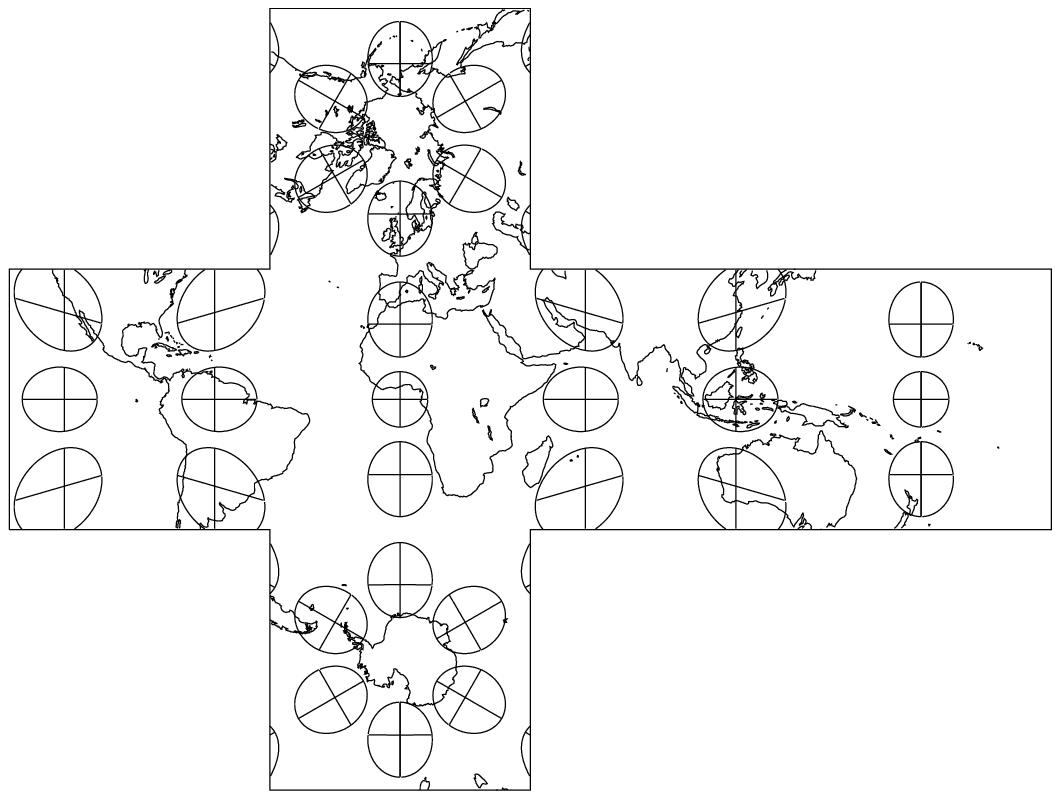,angle=0,height=3.7in}}
\caption{The Indicatrix map for a gnomonic cube projection.}
\end{figure}

\newpage
\begin{figure}[h]
\centerline{\psfig{figure=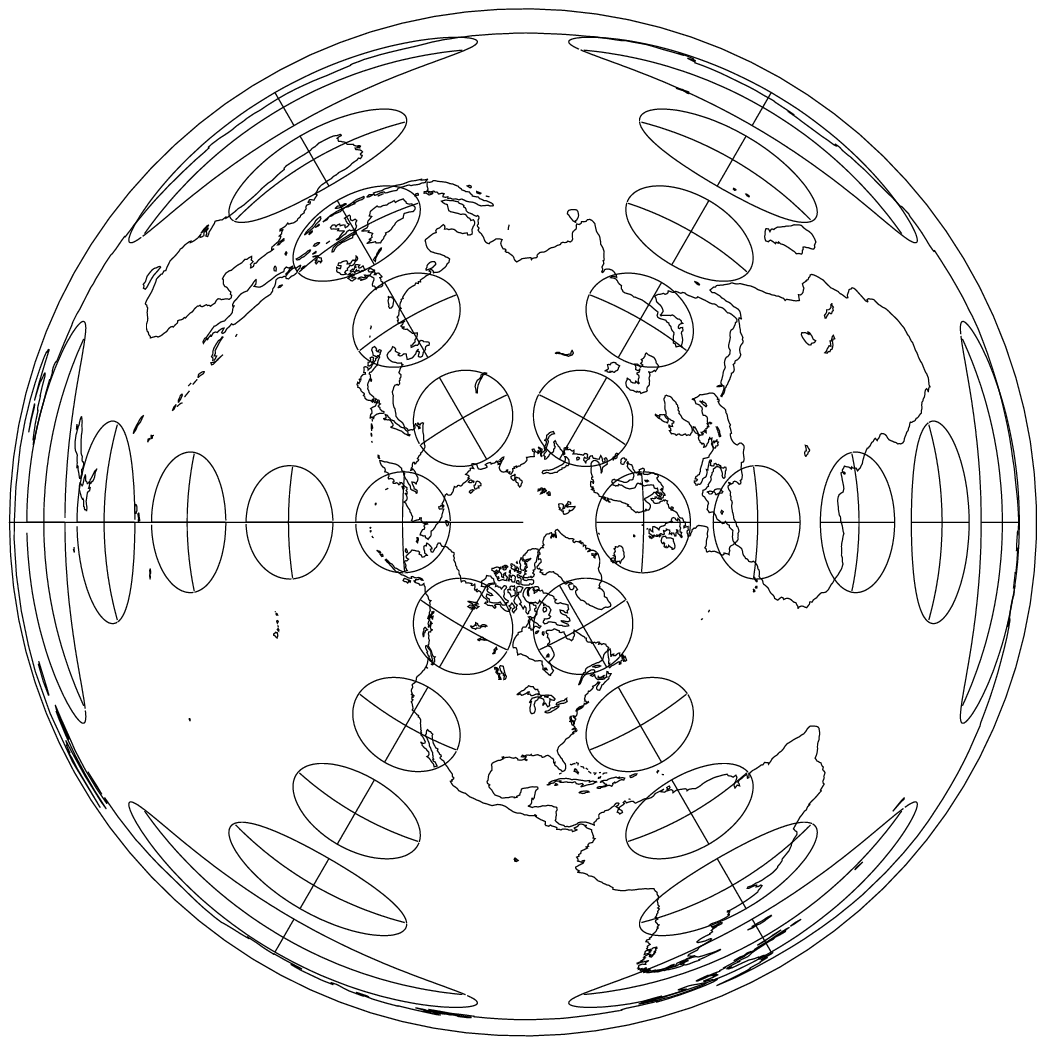,angle=0,height=3.7in}}
\caption{The Indicatrix map for a Gott-Mugnolo projection.}
\vspace{0.2in}
\centerline{\psfig{figure=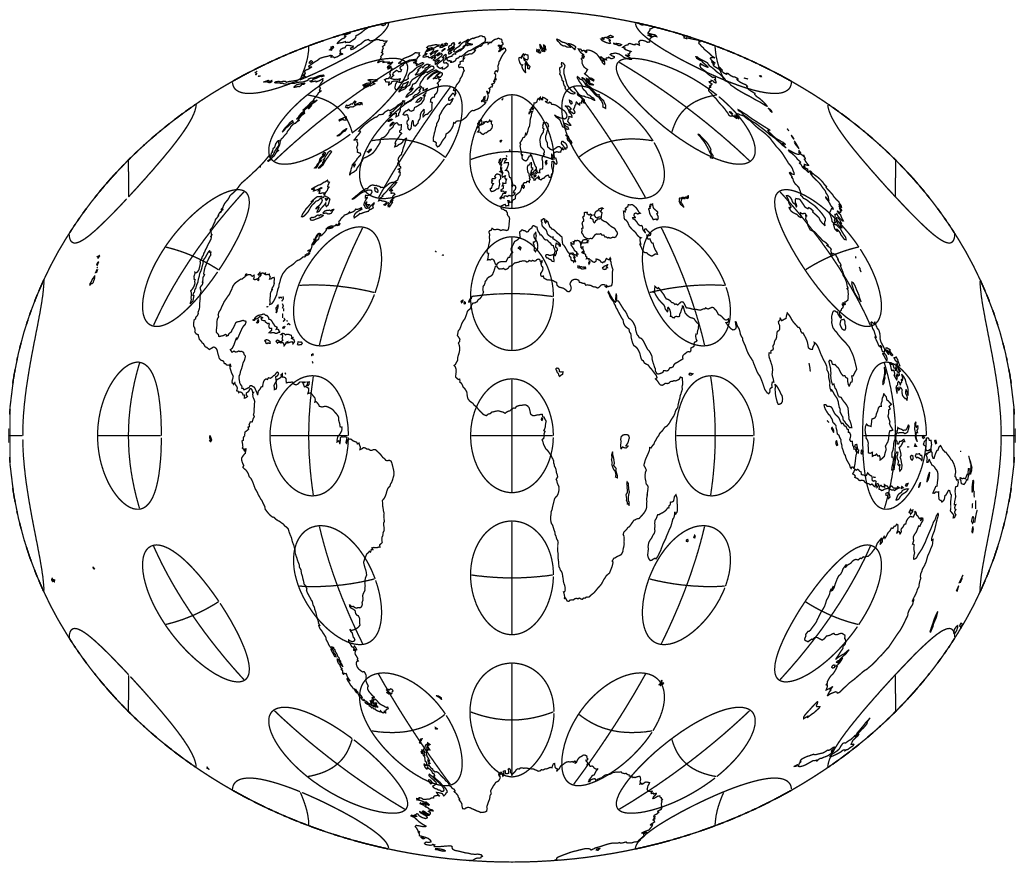,angle=0,height=3.7in}}
\caption{The Indicatrix map for a Gott-Mugnolo Elliptical projection.}
\end{figure}

\newpage
\begin{figure}[h]
\centerline{\psfig{figure=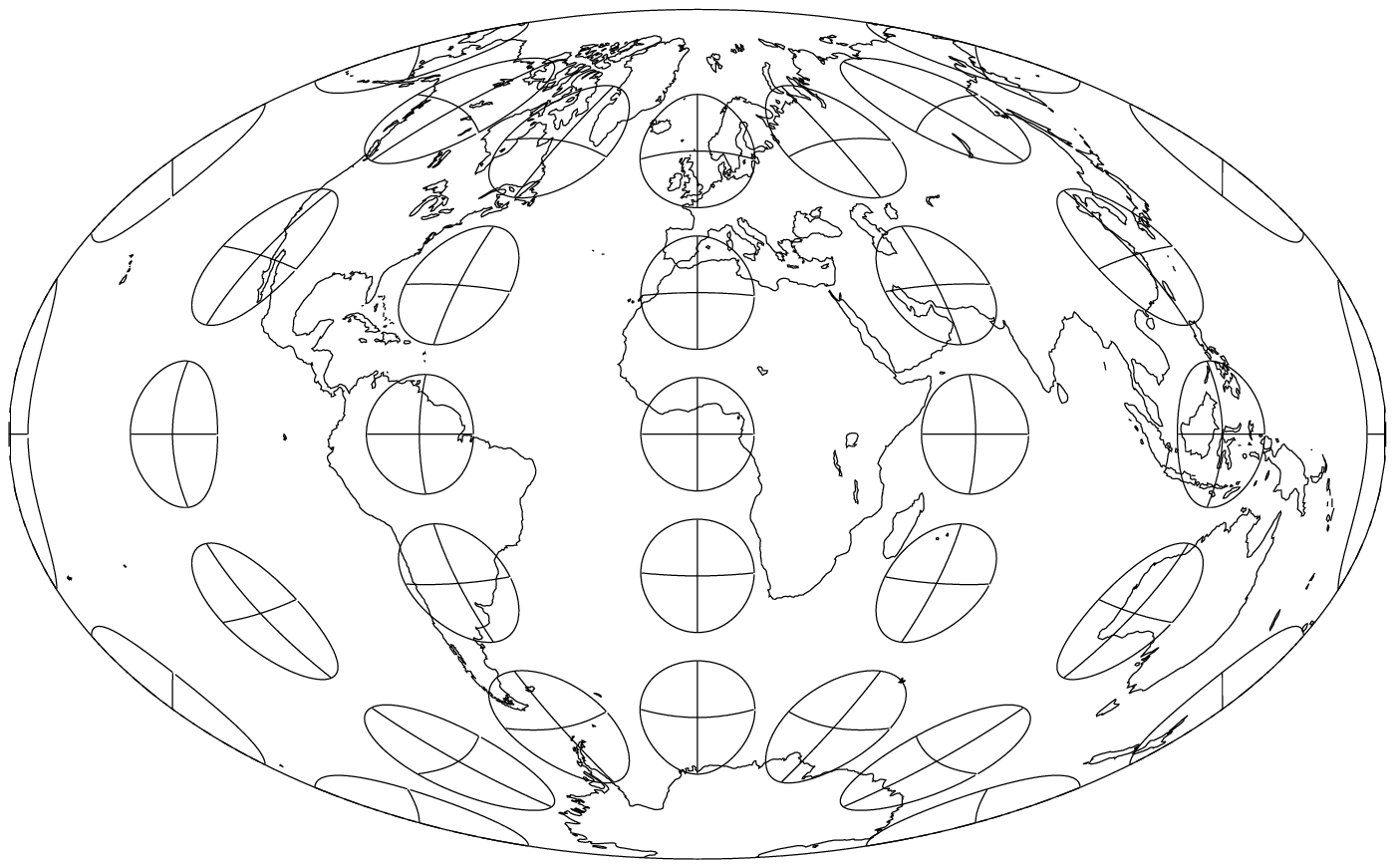,angle=0,height=3.7in}}
\caption{The Indicatrix map for a Gott Equal-Area Elliptical projection.}
\vspace{0.2in}
\centerline{\psfig{figure=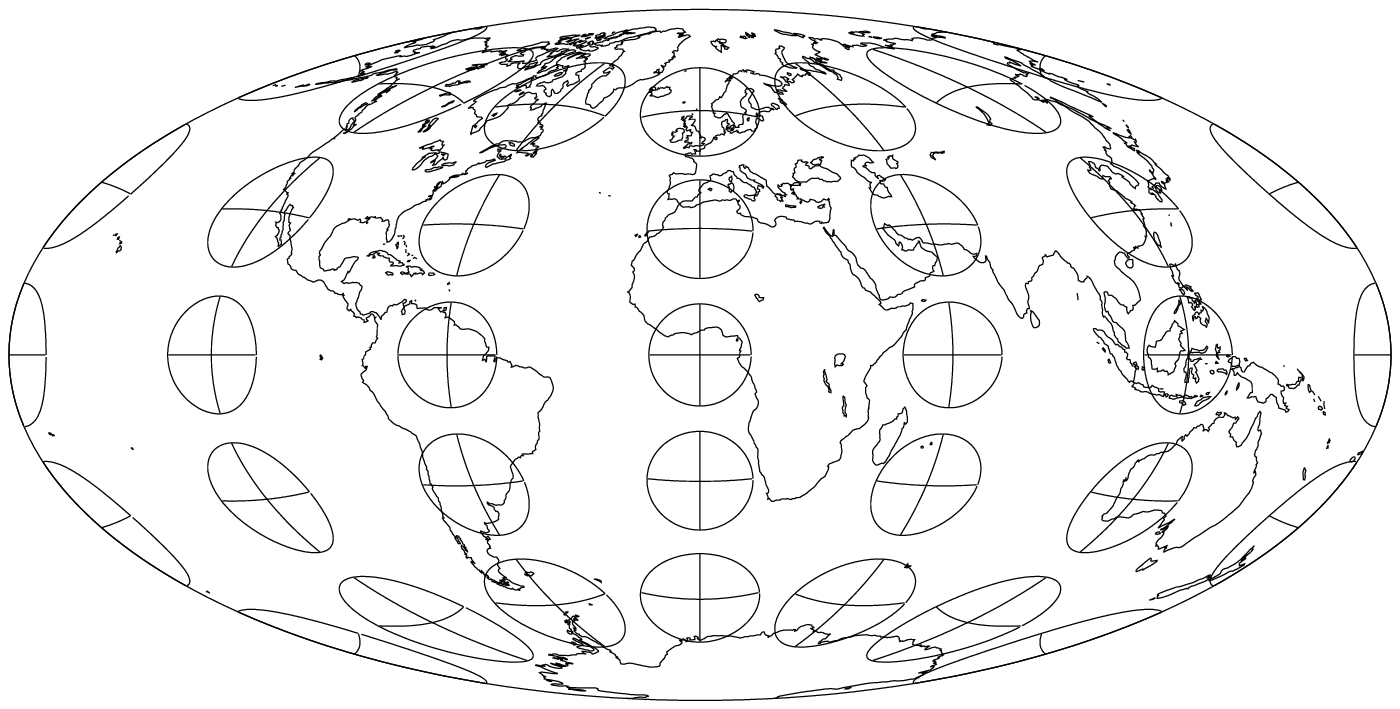,angle=0,height=3.7in}}
\caption{The Indicatrix map for a Hammer projection.}
\end{figure}

\newpage
\begin{figure}[h]
\centerline{\psfig{figure=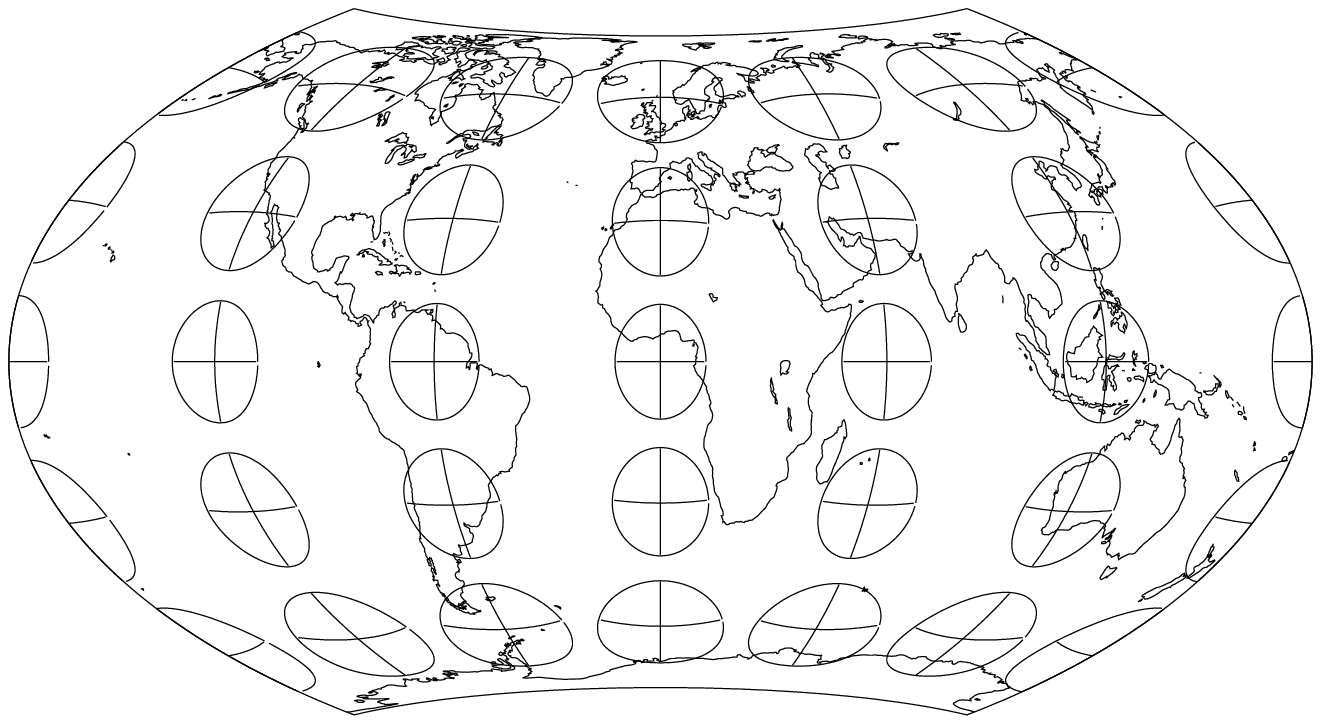,angle=0,height=3.7in}}
\caption{The Indicatrix map for a Hammer-Wagner projection.}
\end{figure}

\newpage
\begin{figure}[h]
\centerline{\psfig{figure=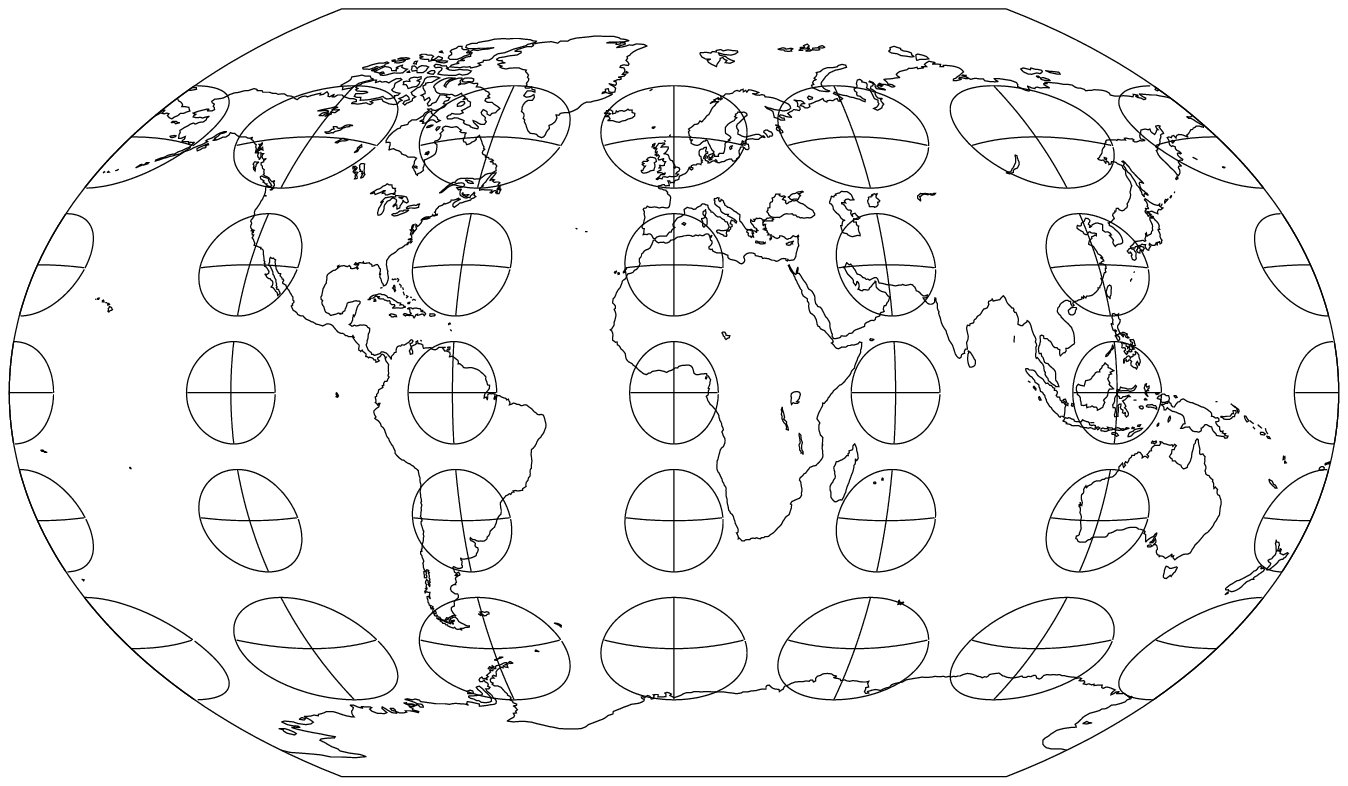,angle=0,height=3.7in}}
\caption{The Indicatrix map for a Kavrayskiy VII projection.}
\vspace{0.2in}
\centerline{\psfig{figure=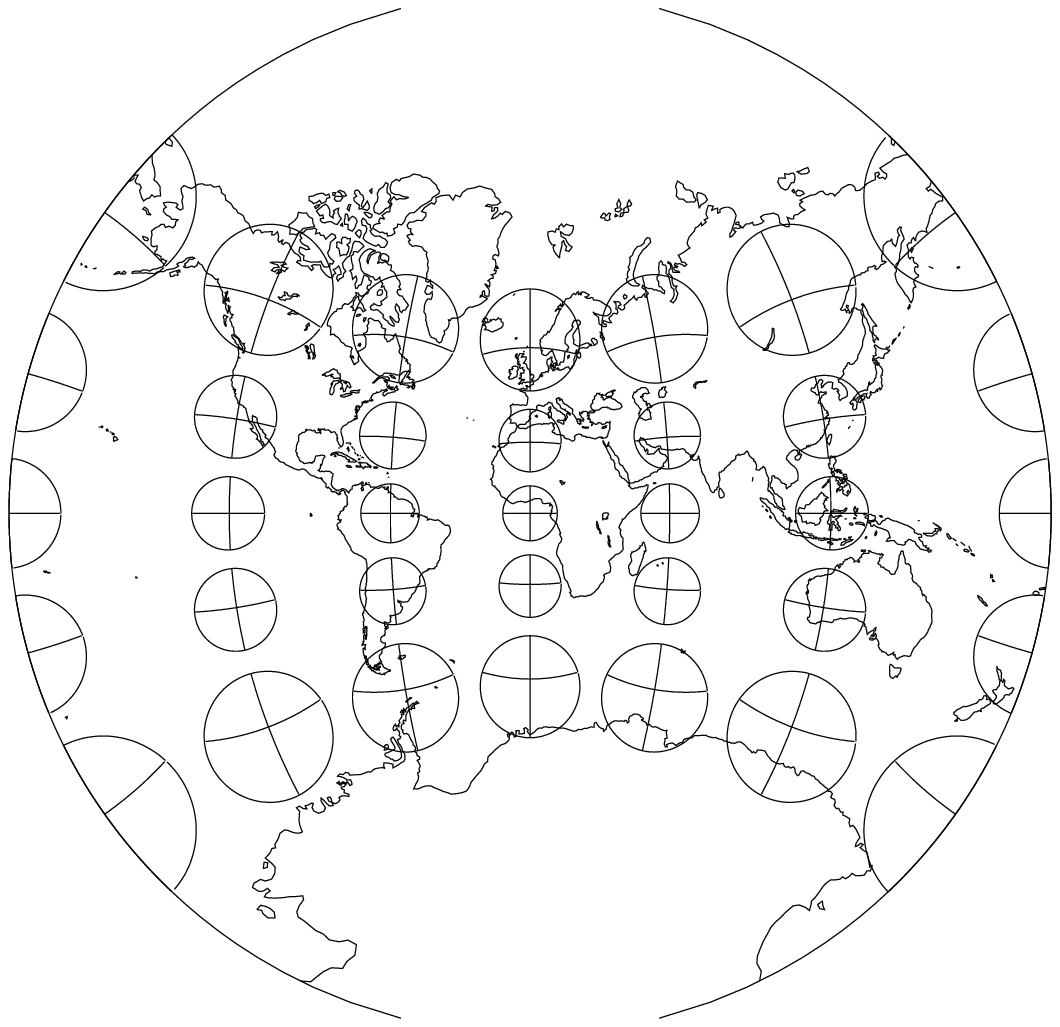,angle=0,height=3.7in}}
\caption{The Indicatrix map for a Lagrange projection.}
\end{figure}

\newpage

\begin{figure}[h]
\centerline{\psfig{figure=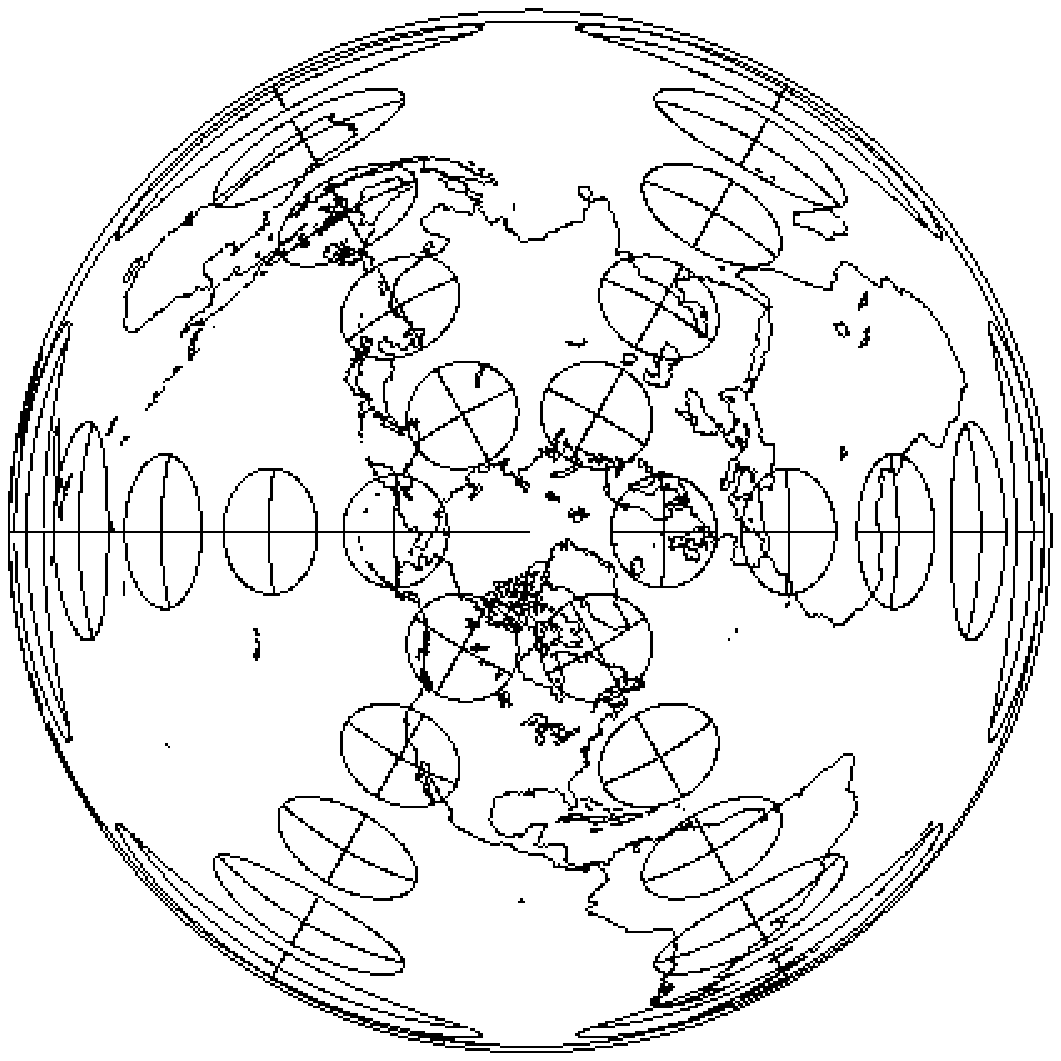,angle=0,height=3.7in}}
\caption{The Indicatrix map for a Lambert Azimuthal projection.}
\vspace{0.2in}
\centerline{\psfig{figure=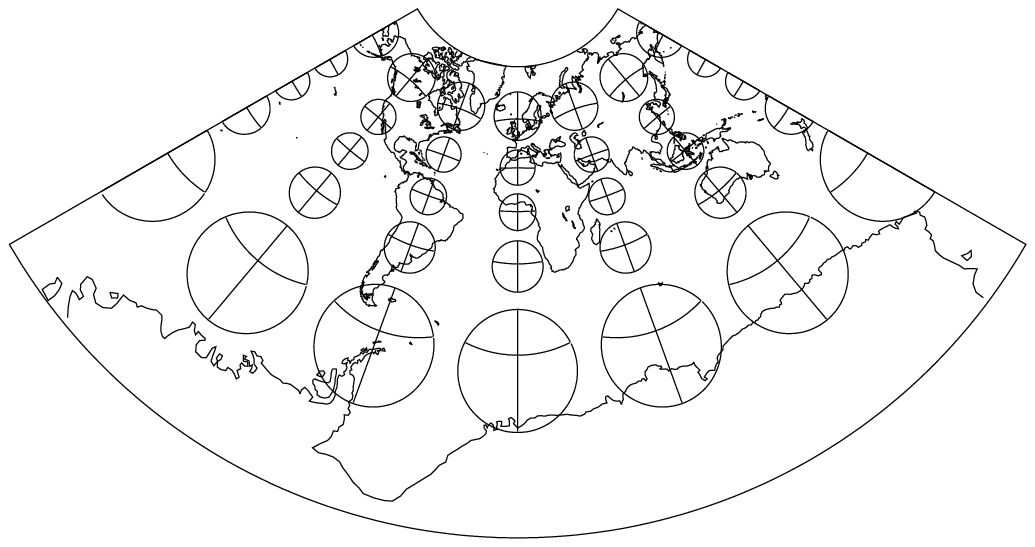,angle=-90,height=3.7in}}
\caption{The Indicatrix map for a Lambert Conic projection.}
\end{figure}

\newpage

\begin{figure}[h]
\centerline{\psfig{figure=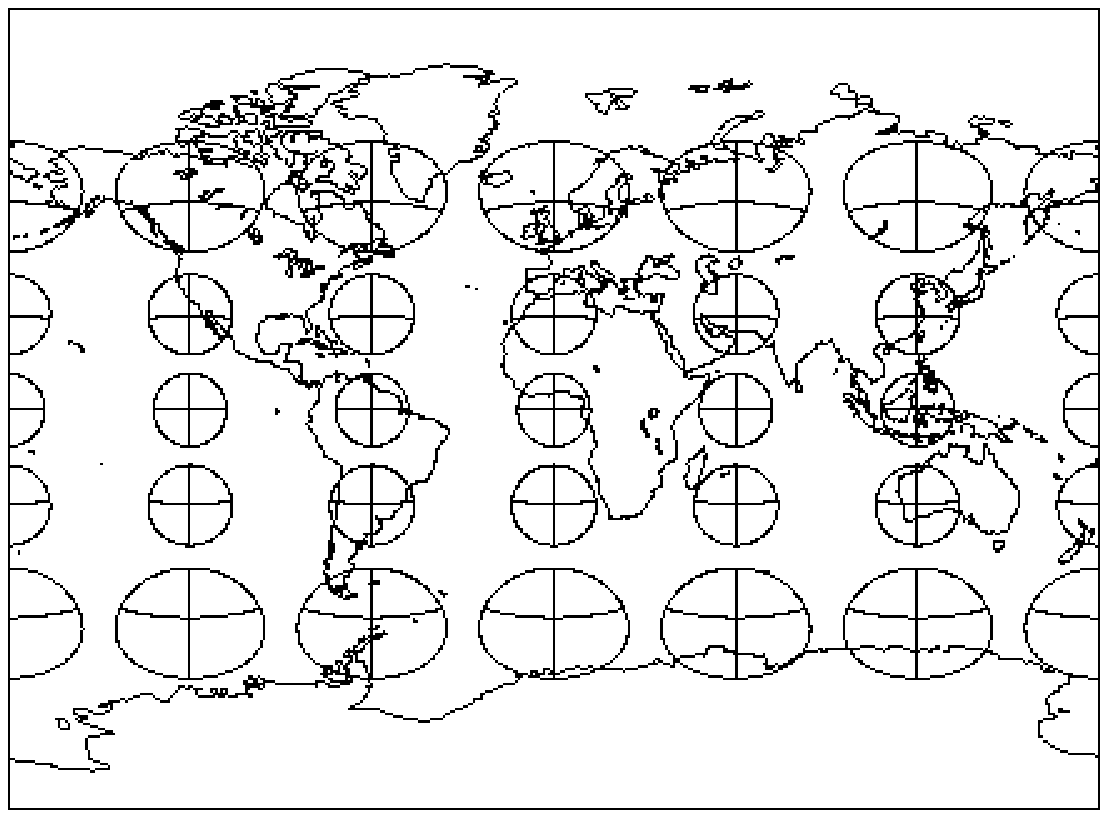,angle=0,height=3.7in}}
\caption{The Indicatrix map for a Miller projection.}
\vspace{0.2in}
\centerline{\psfig{figure=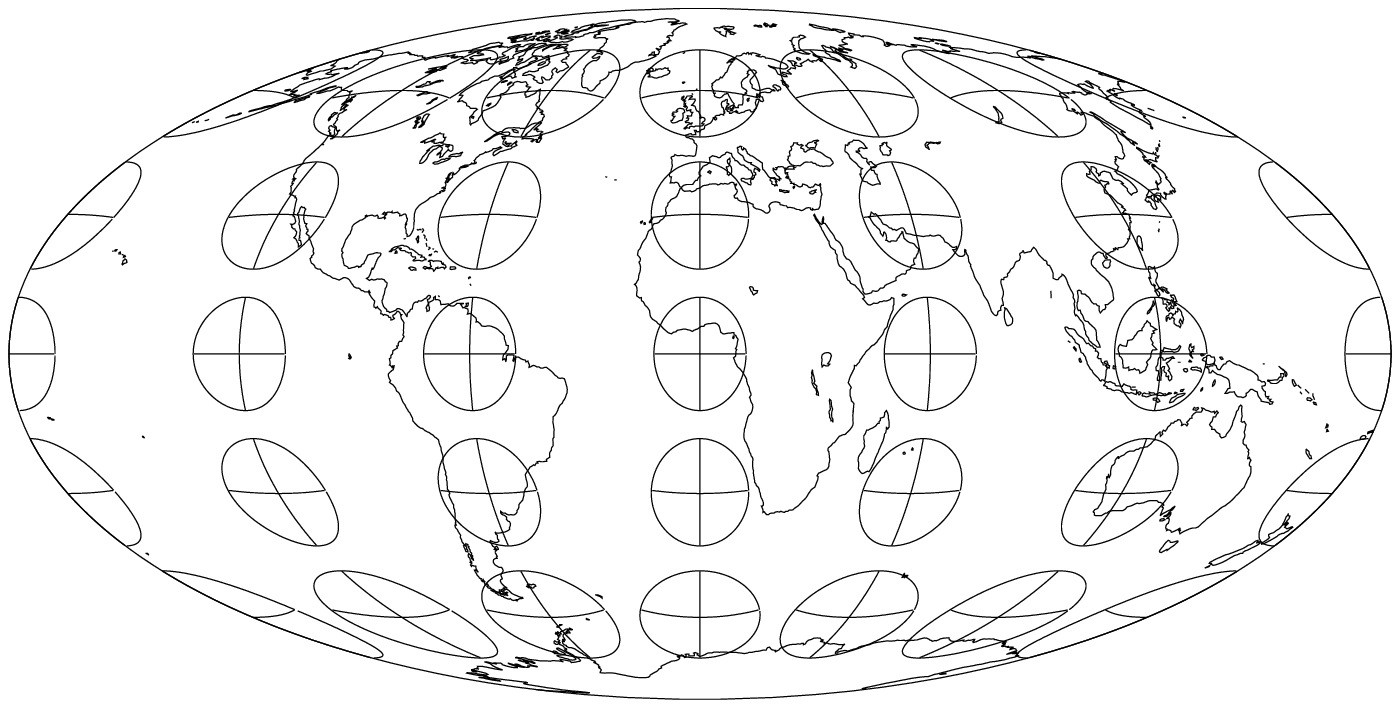,angle=0,height=3.7in}}
\caption{The Indicatrix map for a Mollweide projection.}
\end{figure}

\newpage

\begin{figure}[h]
\centerline{\psfig{figure=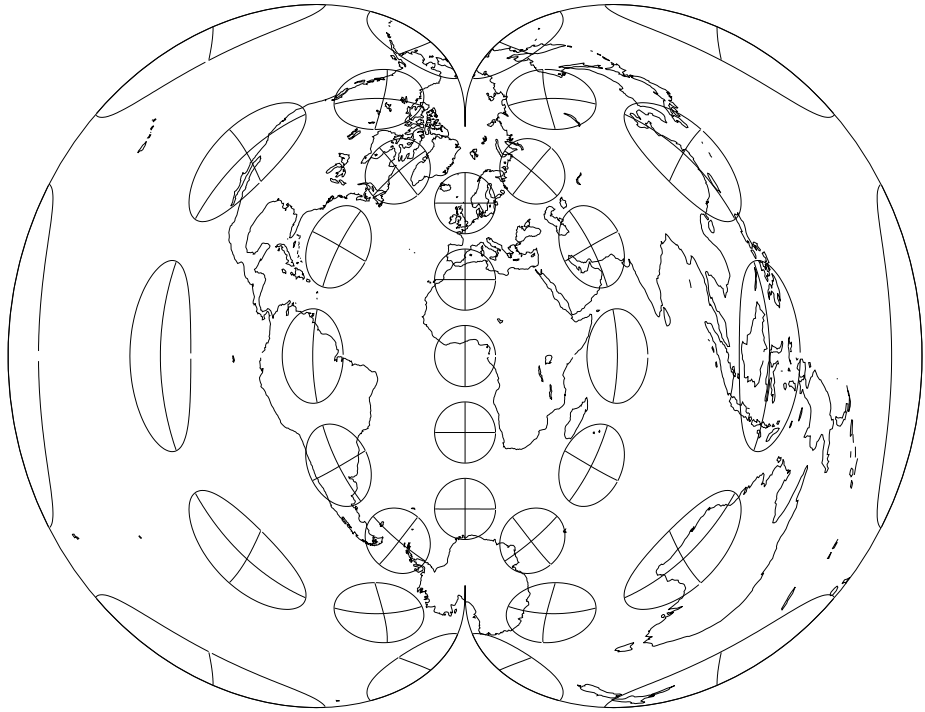,angle=0,height=3.7in}}
\caption{The Indicatrix map for a Polyconic projection.}
\end{figure}

\begin{figure}[h]
\centerline{\psfig{figure=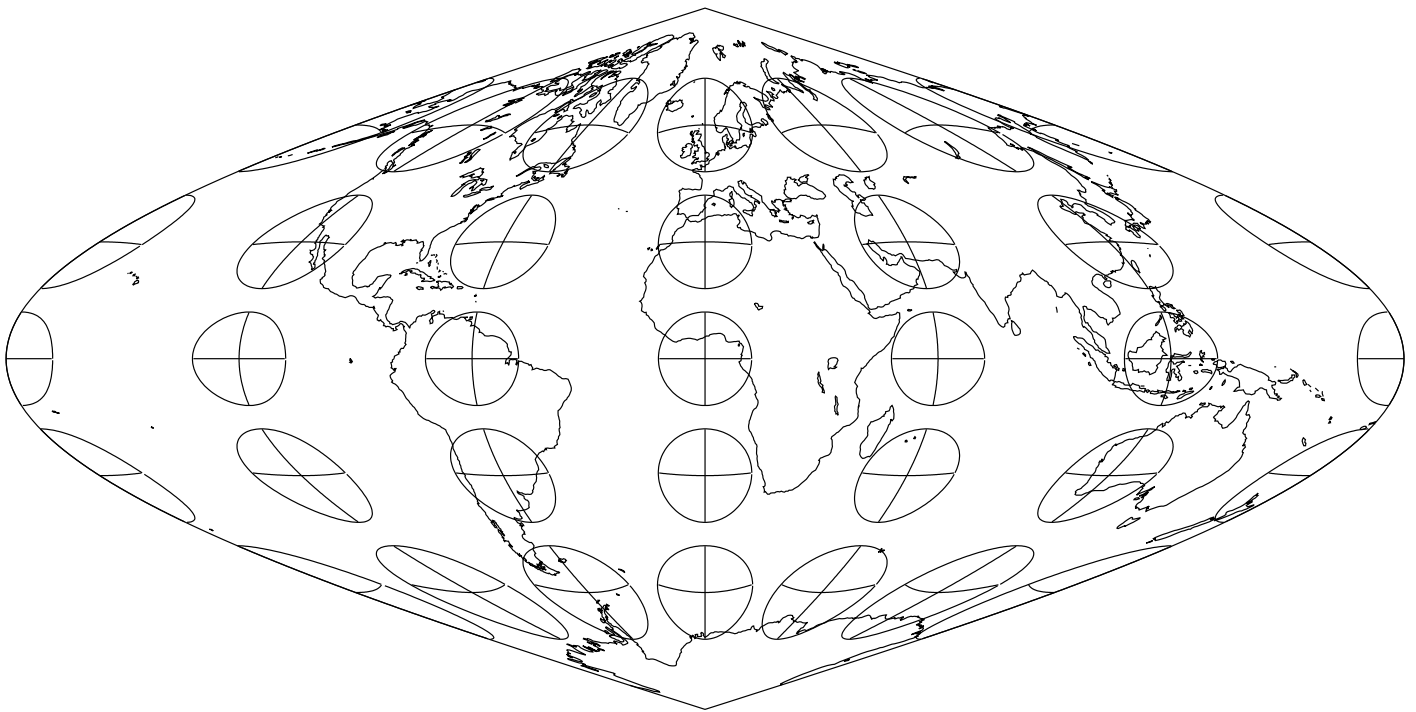,angle=0,height=3.7in}}
\caption{The Indicatrix map for a Sinusoidal projection.}
\end{figure}

\begin{figure}[h]
\centerline{\psfig{figure=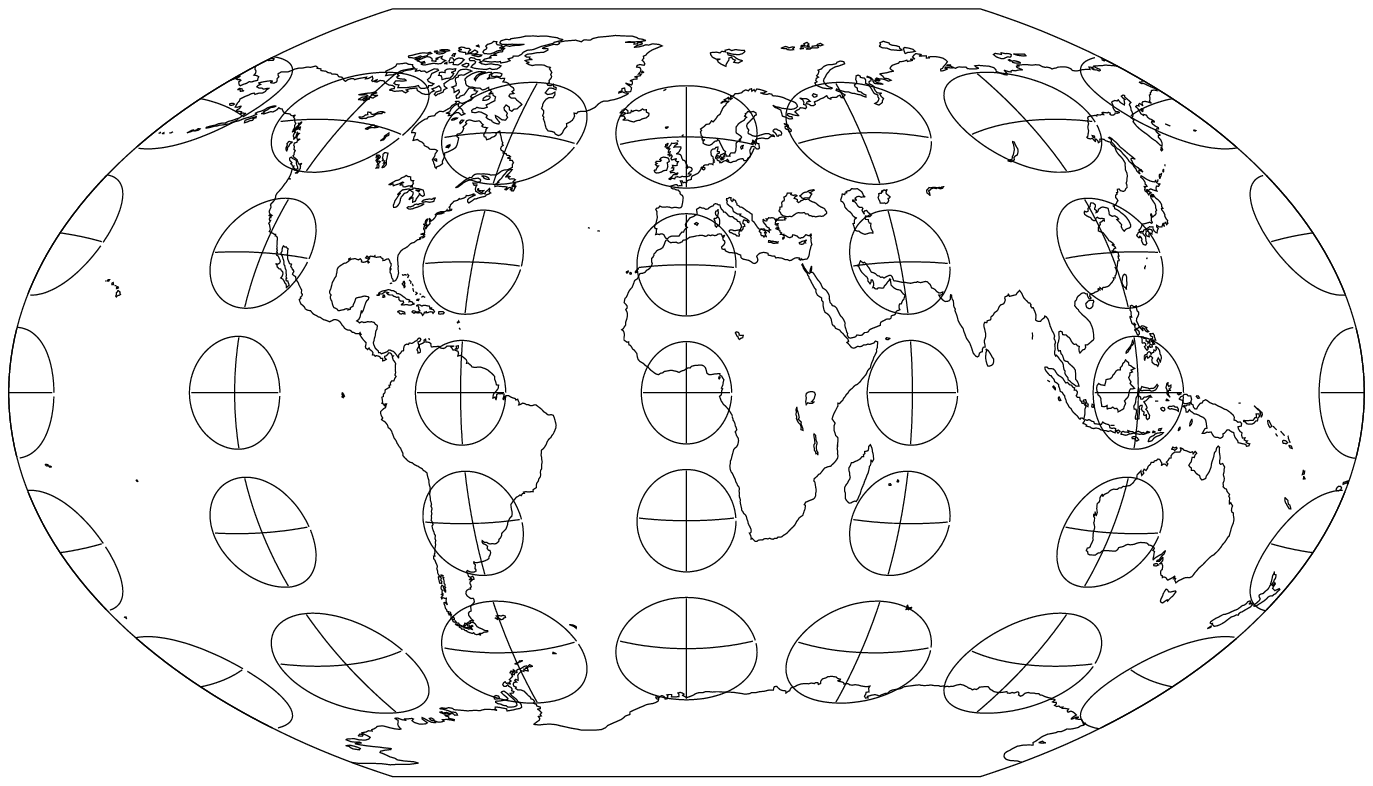,angle=0,height=3.7in}}
\caption{The Indicatrix map for a Winkel-Tripel 
projection.}
\end{figure}

\end{document}